\documentstyle[pra,preprint,aps,epsf]{revtex}
\begin{document}
\draft
\tighten
\title{Prospects for Detecting Supernova Neutrino Flavor Oscillations} 
\author{George M. Fuller,$^{1}$ 
and Wick C. Haxton,$^{2}$ and Gail C. McLaughlin$^{2}$\thanks{Current Address: TRIUMF, 4004 Wesbrook Mall, Vancouver, B.C. Canada V6T2A3.  Electronic address: {\tt gail@alph01.triumf.ca}} }
\address{
$^{1}$Department of Physics, University of California, San Diego,
La Jolla, CA, 92093-0319\\
$^{2}$Institute for Nuclear Theory, Box 351550, and Department of
Physics, Box 351560, \\ 
University of Washington, Seattle, WA 98195, USA\\
}
\date{\today}
\maketitle

\begin{abstract}
The neutrinos from a Type II supernova provide perhaps our best
opportunity to probe cosmologically interesting muon and/or tauon
neutrino masses.  This is because
 matter enhanced neutrino oscillations can lead
to an anomalously hot $\nu_e$ spectrum, and thus to 
enhanced charged current cross sections in terrestrial detectors.
Two recently proposed supernova neutrino observatories,
OMNIS and LAND, will detect neutrons spalled from target nuclei
by neutral and charged current neutrino interactions.
As this signal is not flavor specific, it is not immediately
clear whether a convincing neutrino oscillation signal can 
be extracted from such experiments.  To address this issue
we examine the responses of a series of possible light and
heavy mass targets, $^9$Be, $^{23}$Na, $^{35}$Cl, and 
$^{208}$Pb.  We find that strategies   
 for detecting oscillations which use only neutron count rates
are problematic at best, even if cross 
sections are determined by ancillary experiments. Plausible
uncertainties in supernova neutrino spectra tend to 
obscure rate enhancements due to oscillations.  However, 
in the case of $^{208}$Pb, a signal emerges that is largely
flavor specific and extraordinarily sensitive to the $\nu_e$
temperature, the emission of two neutrons.  This signal
and its flavor specificity are associated with the strength
and location of the first-forbidden responses for neutral
and charge current reactions, aspects of the $^{208}$Pb
neutrino cross section that have not been discussed previously.  
Hadronic spin transfer experiments might be helpful in 
confirming some of the nuclear structure physics underlying
our conclusions.

\end{abstract}
\newpage

\section{Introduction}
\label{sec:intro}
In this paper we investigate some of the difficulties in detecting
the effects of neutrino flavor oscillations on the neutrino 
spectra from Type II supernovae.  In particular,
we examine what might be learned from different target materials
in proposed, long-duration neutrino experiments such as the
Observatory for Multiflavor Neutrinos from Supernovae
(OMNIS \cite{cline91,cline94}) and the Lead Astronomical Neutrino
Detector (LAND \cite{hargrove}).  These detectors would record
neutrons spalled from nuclei following inelastic neutrino excitations.
While neutrons can be produced in either neutral or charge current
interactions, the relative strength of these two contributions 
is sensitive to target thresholds and charge, and thus can be 
adjusted through the choice of target material.  

In this way 
sensitivity to neutrino flavor can be achieved.
For example, one expects a target with low Z and a high charged-current
threshold to be characterized by a low $(\nu_e$,e$^-$) cross section,
and thus to produce neutrons primarily through neutral current
interactions, particularly if the target is also characterized 
by a low neutron separation threshold.  Alternatively, a target
with a high Z, so that Coulomb effects enhance the phase space 
for emitted electrons, and low $(\nu_e$,e$^-$) threshold 
should have a much stronger response to charge current interactions.
The main purpose of this study is to explore what can be
achieved with such target strategies, taking into account the
considerable uncertainties that exist in our understanding of
supernova $\nu_e$, $\bar{\nu}_e$ and heavy-flavor neutrino spectra.

An observation of neutrino flavor transformation, or the demonstration
that this phenomena does not occur over some range of neutrino 
masses and mixing angles, would have
important consequences for both particle physics and astrophysics
(for a review see Ref. \cite{balantekin}).
Neutrino flavor oscillations arise in extended models in which
neutrinos are massive or have magnetic moments, and in which 
the flavor and mass eigenstates are not coincident.
The strength of the flavor mixing can be greatly enhanced in
matter, with two familiar examples being spin-flavor precession
\cite{lim} and the Mikheyev-Smirnov-Wolfenstein
(MSW) \cite{msw} mechanism, with the latter being the most popular proposed
solution of the solar neutrino problem.  

The deficit of solar
neutrinos relative to the predictions of the standard solar model
can be explained by $\nu_e \rightarrow \nu_\mu$ or
$\nu_e \rightarrow \nu_\tau$ flavor oscillations (or by an oscillation
to a sterile state $\nu_e \rightarrow \nu_s$).
The favored MSW solution for the sun suggests that
the mass-squared difference between $\nu_e$ and the second neutrino
involved in the oscillation is
$ \delta$m$^2 \sim 10^{-5}$ eV.  If this second neutrino
is the $\nu_\mu$, then the seesaw
mechanism \cite{seesaw} predicts a mass hierarchy where the $\nu_\mu$
mass $\sim$ few $\times 10^{-3}$ eV and the $\nu_\tau$ mass
is in or near the cosmologically interesting range,
1 to 100 eV \cite{babu}.  This is an attractive scenario as
it allows the $\nu_\tau$ to be a source of hot dark matter.

If neutrino oscillations are responsible for the solar neutrino problem,
similar effects should arise for supernova neutrinos.
Very general arguments lead to a hierarchy of 
average energies for supernova neutrinos, 
 $\langle E_{\nu_\tau}\rangle \sim \langle
E_{\bar\nu_\tau}\rangle
\sim \langle E_{\nu_\mu}\rangle \sim \langle
E_{\bar\nu_\mu}\rangle >
\langle E_{\bar\nu_e}\rangle > \langle E_{\nu_e}\rangle$.
This pattern is established near the neutrinosphere 
(roughly the surface of the neutron star),
where the neutrinos decouple from the matter at a density
of $\sim 10^{12}$ g cm$^{-3}$.

Neutrino oscillations can alter this pattern in a distinctive way,
producing a characteristic signature in terrestrial supernova
detectors, given an MSW neutrino mass level 
crossing outside the neutrinosphere.
As the density at the
neutrinosphere is 10 orders of magnitude greater than that of
the solar core, such crossings occur for an extended range
of $\delta m^2 = m_H^2 - m_L^2$, where $m_H$ and
$m_L$ are the masses of the heavy and light neutrino eigenstates
being mixed.  The resulting values,
$10^{-5}$ eV$^2$ $\leq$ $\delta$ m$^2$ $\leq$ $10^6$ eV$^2$,   
encompasses not only the MSW solutions discussed
in connection with the solar neutrino problem, but also mixing that
might be associated with cosmologically interesting tauon neutrino
masses.

Neutrino flavor transformation can also have important consequences
for
supernova dynamics and nucleosynthesis.  After collapse and core
bounce, the
energy spectra
of neutrinos emitted from the neutrino sphere of the cooling
proto-neutron star
are approximately
Fermi-Dirac, with small chemical potentials. Although a crude
equipartition of
energy between neutrino species is imposed by the weak equilibrium
that obtains in the core, the subsequent decoupling of the 
neutrinos from the matter at the neutrinosphere is flavor 
dependent and leads to the hierarchy of average energies noted
above. The $\nu_\mu$,
$\bar{\nu}_{\mu}$, $\nu_\tau$, and
$\bar{\nu}_\tau$ species decouple deepest in the core because they lack
charged current reactions with nucleons and have smaller cross
sections for scattering off electrons than the $\nu_e$ and $\bar \nu_e$
species.
The $\nu_e$s have the lowest average energy because they are 
the last to decouple: matter near
the neutrinosphere is partially deleptonized and thus rich in
neutrons, enhancing $\nu_e$ + n $\rightarrow$ p + e$^-$.
For example, in one study the
 $\nu_\mu$, $\bar{\nu}_{\mu}$, $\nu_\tau$, and
$\bar{\nu}_\tau$ have average energies $\langle E_{\nu_\mu}
\rangle \sim 25 \, {\rm MeV}$, while the electron neutrinos and
antineutrinos have energies $\langle E_{\bar{\nu}_e}
\rangle \sim 16 \,  {\rm MeV}$ and $\langle E_{\nu_e}
\rangle \sim 11 \, {\rm MeV}$ \cite{qian}.
Neutrinos may be responsible for the revival of the supernova
shock wave, which stalls in most numerical simulations at
a radius of around
200 - 400 km above the surface of
the protoneutron star shortly after core bounce, $t_{\rm pb}
\sim$ 0.1 s.  Neutrino interactions in the nucleon gas left in the
wake of the shock wave can deposit considerable energy, providing
the push needed for a successful explosion.  Oscillations can
enhance this effect:
If a $\nu_e \leftrightarrow \nu_\tau$
oscillation took place between the edge of the neutron star and the
stalled shock at this epoch, the resulting more energetic $\nu_e$
flux increases the rate of neutrino heating \cite{fuller92}.
Neutrino flavor oscillations can also alter 
supernova nucleosynthesis at later times $t_{\rm pb} {\
\lower-1.2pt\vbox{\hbox{\rlap{$>$}\lower5pt\vbox{\hbox{$\sim$}}}}\ }
3\,{\rm
s}$ \cite{qian}.

Terrestrial experiments exploiting accelerator or reactor neutrino sources, such as LAMPF's
Liquid Scintillator Neutrino Detector (LSND \cite{lsnd})
and the CERN experiments NOMAD \cite{nomad} and CHORUS
\cite{chorus}, are placing constraints on vacuum oscillations.
To date, no evidence for $\nu_\mu \leftrightarrow
\nu_\tau$ mixing has been found at NOMAD or CHORUS.  LSND has attributed an
excess of events above background to
$\bar{\nu}_\mu \rightarrow \bar{\nu}_e$, although
these events have also been interpreted as an upper limit \cite{lsnd}.
The KARMEN \cite{karmen} experiment, which is similar to LSND in its sensitivity
to $\delta$m$^2$ and mixing angle, has not yet accumulated enough
data to convincingly confirm or rule out the LSND result.
But perhaps the strongest indication of oscillations comes from 
the deficit of muons in the interactions of upward going 
atmospheric neutrinos, as recently reported by the SuperKamiokande
collaboration \cite{atmos}.  
The solar neutrino problem, atmospheric neutrino anomaly, and 
LSND results all suggest new physics, though all of these 
anomalies together are difficult to
reconcile with a simple pattern of neutrino masses and mixing angles
arising in theories with only 
three active neutrinos \cite{cf}.

Important new constraints on neutrino properties can be extracted
from observations of supernova neutrinos.  One technique for
measuring neutrino mass, independent of mixing angles, exploits
the time delay and/or spreading in the
neutrino signal (see for example Ref. \cite{cline94}). The arrival
time difference
for $\nu_e$ and $\nu_\tau$ neutrinos with masses $m_{\nu_e}$ and
$m_{\nu_\tau}$, respectively, is
\begin{equation}
\delta t \sim 0.514 R_{10 \, {\rm kpc}}
[ (m_{\nu_\tau} / E_{\nu_\tau})^2 - (m_{\nu_e} / E_{\nu_e})^2],
\end{equation}
where $ E_{\nu_\tau}$ and $E_{\nu_e}$ are the energies of the
tauon and electron neutrinos, $R_{10 \, {\rm kpc}}$ is the distance
to the supernova in 10 kiloparsecs (comparable to the galactic
radius), and $\delta t$ is measured in seconds.
(Alternatively, one can rewrite Eq. (1) for a single flavor,
but with arrival times dependent on the neutrino energy.)
The result is a characteristic spreading of the neutrino pulse,
with arrival times correlated with the neutrino energy and/or flavor.
Neutrino masses, or limits on masses, can
be deduced by comparing an observed neutrino signal with 
the spectra and time-dependent luminosities arising in plausible
supernova models.  

Measurements made by Kamiokande and IMB at
the time of SN1987A were argued to provide a limit on the $\bar \nu_e$ mass.
The analysis were limited by the small number of detected neutrino events
and by uncertainties in modeling the supernova mechanism and 
associated neutrino emission \cite{mayle}.  As a result, the deduced
limits span a considerable range.  Clearly such astrophysical 
uncertainties will also affect future time-of-flight neutrino
mass limits derived from new detectors like OMNIS and LAND.  Yet
these detectors should have two important advantages.
First, they promise a large number of neutrino
events for a galactic supernova, possibly giving us a detailed time history 
of neutrino emission associated with the supernova.  For example, 
it was recently argued that large event rates would allow 
experimentalists to map out the expected initial sharp rise in neutrino
emission following core bounce, a feature in the neutrino cooling
curve that could be exploited to significantly tighten
mass limits \cite{totani}.  Second, complimentary
information from other new detectors, such as SuperKamiokande \cite{sk},
will reduce the degree to which analyses must depend on 
poorly understood aspects of supernova models.  The spectrum and flavor
of supernova neutrinos will be more accurately characterized given
a complement of detectors with different thresholds and flavor
sensitivities.  Flavor specificity in time-of-flight measurements
is quite important because competing laboratory
limits on the $\nu_\tau$ and $\nu_\mu$ masses, 24 MeV and 170
keV, respectively, are so poor.

If neutrinos mix, supernovae could provide an important consistency check 
on models of neutrino masses and also possibly on time of flight derived 
neutrino masses.  Flavor oscillations, enhanced by matter
effects, can lead to transformation between $\nu_e$'s and either
the $\nu_\mu$ or $\nu_\tau$, leading to an anomalously energetic
$\nu_e$ spectrum.  This departure from the usual hierarchy of 
average neutrino energies is a powerful test for new physics 
because it will occur for an extended range of $\delta$m$^2$
and mixing angles. In fact, the neutrino mass level crossings become 
increasingly
adiabatic for larger $\delta$m$^2$, with adiabatic flavor transformation
occurring for mixing angles sin$^2 \theta \gtrsim 10^{-5}$.  Thus the 
observation
of an excess of supernova $\nu_e$ events provides an opportunity to
probe neutrino phenomena that may be inaccessible otherwise.

Several detectors, both in operation and proposed, 
could detect neutrinos from a galactic supernova.   (A partial
review can be found in \cite{burrows}.)  
Two of particular note are the light water Cerenkov detector
SuperKamiokande, which has an inner fiducial volume
of 22.5 kilotons and has been in operation for approximately 
two years; and the Sudbury Neutrino Observatory \cite{SNO},
a heavy water Cerenkov detector whose inner vessel will contain
one kiloton of D$_2$O.  SNO is currently in its commissioning phase
and should be fully operational by the end of 1998.
In SNO charged and neutral current reactions will produce 
distinct signals.  The neutral current neutrino reaction
D$(\nu_{\rm x},\nu'_{\rm x})$np produces free neutrons.
These will be detected either by their (n,$\gamma$)
reactions on $^{35}$Cl, which will be introduced by dissolving
salt in the water, or by their interactions in specially designed
counters utilizing the $^3$He(n,p) reaction.  The charged
current reaction D($\nu_e$,e$^-$)pp produces energetic electrons 
that will be observed through Cerenkov light.  (The absence
of coincident neutrons distinguishes this reaction from 
D($\bar \nu_e$,e$^+$)nn.)  A supernova neutrino burst altered
by $\nu_e \leftrightarrow \nu_\mu/\nu_\tau$ oscillations will produce an enhanced ($\nu_e$,e$^-$)
signal, while leaving the rest unchanged.

SuperKamiokande is of particular interest because of its size and
its likely longevity: the collaboration hopes to operate the 
detector for three decades, a period approaching the timescale
for galactic supernova.  However the enormous event rate for
($\bar \nu_e$,e$^+$) off free 
protons tends to obscure, in the case of flavor oscillations, the 
$\nu_e$ signal of interest.  Perhaps the best opportunity for
measuring the $\nu_e$'s is through the reaction $^{16}$O($\nu_e$,e$^-$),
which produces a back-angle enhancement in the electron 
distribution that will distort the known (and nearly isotropic)
distribution from ($\bar\nu_e$,e$^+$) \cite{haxton16o}.  

In contrast, the flavor 
oscillation effects on the forward-peaked events from $\nu$-electron
scattering are very subtle and difficult to extract. This
cross section is approximately linear in the neutrino energy and so
there is no net change in the event rate due to flavor oscillations.
The event rate is then proportional to the luminosity, which we
noted earlier was approximately independent of flavor.  Note that this 
contrasts with semileptonic interactions, where cross sections
scale as E$_\nu^2$ or faster, depending on nuclear thresholds.
Yet there is a shift in the distribution of forward-peaked events 
towards higher energy from neutrino-electron scattering.
This is because the $\nu_e$-electron cross section is
approximately six times that the cross section for heavy flavor neutrinos.
In turn, this effect may provide a signal for flavor oscillations 
\cite{minakata}.  

Another interesting possibility, suggested quite
recently \cite{caltech}, is the detection of the 5-10 MeV $\gamma$ rays
produced in cascades following the neutral current breakup of $^{16}$O.  
A supernova at a distance of 10 kpc would produce a few hundred
such events from $\nu_\mu$ and $\nu_\tau$ interactions in
SuperKamiokande.  A tauon mass could then be extracted from analysis of 
the time evolution of the signal \cite{beacom}.

One of the arguments for detectors such as OMNIS and LAND is that
they could remain in operation over a long period of time, making the
probability of observing a galactic supernova reasonably high.  These
detectors
would record neutrons produced in the neutral
current breakup of nuclei,
\begin{equation}
\label{eq:nc}
\nu_i + ({\rm Z},{\rm N}) \rightarrow  ({\rm Z},{\rm
N-1})
+ \nu_i + n,
\end{equation}
where $i$ represents all neutrino and antineutrino species.  Here
(Z,N) denotes a nucleus with Z protons and N neutrons.  A similar
signal can arise for the analogous charged current reactions
\begin{equation}
\label{eq:cc}
\nu_e + ({\rm Z},{\rm N}) \rightarrow  ({\rm
Z+1},{\rm N-2})
+ e^{-} + n
\end{equation}
and
\begin{equation}
\label{eq:ccb}
\bar \nu_e + ({\rm Z},{\rm N}) \rightarrow  ({\rm
Z-1},{\rm N})
+ e^{+} + n.
\end{equation}

By itself, the observation of a neutron in OMNIS or LAND provides no
information on the type of initiating neutrino reaction.
The goals of this paper include calculating the cross sections and spallation probabilities for these
detectors more carefully than has been attempted before;
exploring to what extent the use of multiple nuclear targets
might enhance flavor sensitivity; and exploring what can be
learned by comparing the rates for one and two neutron spallation.
Ideally one would hope to find targets with very different 
relative sensitivities to $\nu_e$ and neutral current reactions.
The success of such a strategy clearly depends on our ability
to accurately calculate (or measure) the neutrino responses of the targets,
and to estimate uncertainties in supernova flux predictions.

In section \ref{sec:neutphys}
we discuss neutrino-induced neutron spallation
in both  high Z and low Z target materials,
describing the underlying nuclear structure physics governing the
responses.  We also provide estimates of cross sections for
four possible target materials,
$^9$Be, $^{23}$Na, $^{35}$Cl and
$^{208}$Pb.
In section \ref{sec:detect}  
we discuss strategies for determining whether the neutrino flux 
has been altered by oscillations.  Although our study is by no
means exhaustive, it appears that the tactic of looking
for changes in total spallation cross sections is rather
challenging.  It is very difficult, even using multiple targets, to
achieve the necessary degree of sensitivity to the $\nu_e$ temperature. 
The primary difficultly is our uncertain knowledge of the spectrum
of supernova neutrinos in the absence of oscillations.
The one exception we found to this general rule is the two
neutron spallation channel in $^{208}$Pb, which appear to 
provide an exquisitely sensitive $\nu_e$ thermometer.  The
underlying physics involves the first-forbidden contributions 
to the charged and neutral current channels which have not
been considered previously.
We suggest some experimental work that would help in characterizing
the $^{208}$Pb response to neutrinos.    

\section{Neutrino-Nucleus Interactions}
\label{sec:neutphys}
In this section we discuss supernova neutrino reactions with
nuclear targets which lead to the spallation of one or more neutrons.
There are three main physics issues.  The first is estimating
the target response: what is the distribution of final nuclear
states that will result when target nuclei interact with an
incident spectrum of neutrinos?  For the relatively low neutrino
energies of interest, the nuclear response is dominated by 
allowed and first-forbidden transitions.  
Fortunately we have a number of experimental tests of these
responses, and there exist approximate sum rules that are both
important guides to and constraints on calculations. 

The second issue is the probability that a neutrino interaction
will result in the emission of a neutron, thus producing a
signal in the detector.  Neutron emission can only occur if 
the daughter nucleus is excited above the neutron separation
energy.  The branching ratio into this channel also depends on
the competition with other open channels, such as proton
or $\alpha$ emission.  We estimate these in Hauser-Feshbach 
calculations.

The third issue is the supernova neutrino spectrum.  Because the 
threshold for neutron spallation can be substantially, often the
high energy tail of the neutrino spectrum is especially important
in determining the overall rate.  Various numerical simulations
of supernova explosions differ in the approximations made
in treating neutrino diffusion, convection, etc.  Thus, while
there is qualitative agreement about the average energy 
hierarchy discussed in the introduction, there are differences
in the precise value of the average energy and in the details
of the spectrum shape.  The resulting uncertainties clearly
have an influence on predictions of flux-averaged nuclear cross
sections.

The last of these issues, the neutrino spectrum, enters in evaluating
the flux-averaged cross section
\begin{equation}
\label{eq:cross}
\langle \sigma \rangle = {\int_{E_{th}}^{\infty} f_\nu(E_\nu)
\, \sigma(E_\nu) dE_\nu},
\end{equation}
where $E_{th}$ is the threshold energy for the reaction,
$f_\nu$ is the normalized neutrino spectrum,
and $\sigma(E_\nu)$ is the nuclear
cross section for an incident neutrino of energy $E_\nu$.
The supernova
neutrino energy spectra predicted by transport codes can be
represented approximately by modified
Fermi-Dirac distributions of the form \cite{wilson,janka}
\begin{equation}
f_\nu = \left[ {{1}\over{T_{\nu}^3 F_2(\eta_{eff})}}\right]\,
{E_\nu^2
\over \exp(E_{\nu} / T_{\nu} - \eta_{eff} ) +1}.
\end{equation}
Here $T_\nu $ and $\eta_{eff} $ are the neutrino
temperature and
degeneracy parameter (chemical potential divided by $T_\nu $),
respectively,
and $F_2(\eta_{eff})$ is the relativistic Fermi integral of order 2
and
argument $\eta_{eff}$, required to normalize the above distribution
to unity.   The Fermi integrals of order $k$ are defined
by
\begin{equation}
\label{eq:fermi}
F_k(\eta ) \equiv \int_{0}^{\infty}{{x^k dx \over \exp(x - \eta )
+1}}.
\end{equation}
The flux $d \Phi_\nu$ of neutrinos with energies between $E_\nu$
and $E_\nu + dE_\nu$ a large distance $r$ from a supernova can
then be written
\begin{equation}
d\Phi_\nu(E_\nu) =  {{L_\nu}\over{4\pi r^2}}
{{1}\over{\langle E_\nu
\rangle}} f(E_\nu ) dE_\nu ,
\end{equation} where $L_\nu$ is the luminosity of the neutrino
species of interest.  Note that $\langle E_\nu \rangle$ = 
$T_\nu F_3(\eta_{eff})/F_2(\eta_{eff})$ and is $\sim 3.15T_\nu$
when $\eta_{eff}$=0 and $\sim 3.99T_\nu$ when $\eta_{eff}$ = 3.

Predictions of neutrino energy spectra and luminosities vary between
different
supernova neutrino transport codes, thus producing
different values of  $\eta_{eff}$
and $T_\nu$ when approximated as in Eq. (7).  For example, the transport
calculations by Janka yield spectra with $\eta_{eff} \sim$ 3
for all neutrino species \cite{janka}.  While this choice also
produces a good fit to the $\nu_e$ and $\bar \nu_e$ spectra
of Wilson and Mayle \cite{wilson}, their heavy-flavor neutrino
spectra more closely resemble a black-body distribution
($\eta_{eff} \sim$ 0).  Such differences
are an important source of uncertainties in predicting neutron
counting rates in a detector, a point we will return to in
section III.

We now turn to the issue of the neutrino reaction cross sections.
At typical supernova neutrino energies one expects the total
cross section for the charged current reaction $(\nu_e$,e$^-$)
on a parent nucleus of charge Z to be dominated by the
allowed transitions to the isobaric
analog state (IAS) and the Gamow-Teller (GT) resonance states in the
daughter nucleus.  The allowed cross section is 
\begin{equation}
\sigma(E_{\nu_e}) = 
{G^2_F \cos^2\theta_c \over \pi} k_e E_e F(Z+1,E_e) 
[|M_F|^2 + (g_A^{\rm eff})^2 |M_{GT}|^2],
\end{equation}
where $G_F$ is the Fermi constant, $E_e$ and $k_e$ are the energy and
three-momentum of the outgoing electron, respectively,
$\theta_c$ is the Cabibbo 
angle, and $F(Z+1,E_e)$ accounts for the Coulomb distortion
of the outgoing electron wave function, which we take from the
tabulations of Behrends and Janecke \cite{behrends}.  In several cases 
we will study below, the total BGT strength is 
taken from shell model calculations that satisfy
the Ikeda sum rule implicitly (see below).
Phenomenologically it is known that 
these approaches will overestimate low-lying BGT strength unless an effective
axial-vector coupling constant $g_A^{\rm eff} \approx 1$ is used, rather 
than the
bare nucleon value 1.26 \cite{brown}.  Thus we allow for such a 
renormalized $g_A^{\rm eff}$.

The allowed Fermi and GT transition strengths are
\begin{equation}
|M_F|^2={1\over 2J_i+1}|\langle J_f||
\sum_{i=1}^A\tau_+(i)||J_i\rangle|^2,
\end{equation}
and
\begin{equation}
|M_{\rm GT}|^2={1\over 2J_i+1}
|\langle J_f||\sum_{i=1}^A\sigma(i)\tau_+(i)||J_i\rangle|^2,
\end{equation}
respectively.
To evaluate the cross section one must specify the distribution of these
transition probabilities over the 
final states of the daughter nucleus.
All of the formulas above also apply to $(\bar \nu_e$,e$^+)$ 
provided the corresponding Coulomb correction $F(Z-1,E_e)$ is
evaluated for a positron and the isospin operators are replaced
by $\tau_{-}(i)$.
  
In the limit of good isospin the Fermi strength $|M_F|^2 = |N-Z|$
is carried entirely by the IAS in the daughter nucleus.
All of the nuclei of present interest ($^9$Be, $^{23}$Na, $^{35}$Cl,
$^{208}$Pb) are neutron rich, so Fermi transitions contribute only
to the $(\nu_e,$e$^-$) direction.  The Fermi transitions for the 
first three nuclei populate the mirror ground states
of the daughter nuclei, none of which decays by neutron emission.
Thus they are of no interest to us.  The analog state in $^{208}$Bi,
however, is located at an excitation energy of 15.16 MeV, well
above the neutron breakup threshold and just barely (0.2 MeV) above the 
two-neutron breakup threshold.  Therefore
\begin{equation}
|M_F(E)|^2 = 44 \, \delta_{E \, E^\prime},~~ 
^{208}{\rm Pb}(\nu_e,e^-)^{208}{\rm Bi}
\end{equation}
where $E = E_\nu - E_e$ is the nuclear (not atomic) 
excitation energy measured relative to the parent ground state in 
$^{208}$Pb, and $E^\prime \approx 17.53 \, {\rm MeV}$.
  
The GT strength is more complex.  The difference between the GT
strength in the ($\nu_e$,e$^-$) channel and that in the
$(\bar \nu_e$,e$^+$) direction is governed by the Ikeda sum rule,
$\sum_f |M_{\rm GT}|^2 \sim 3(N-Z)$, but this sum rule is
generally not saturated by the low-energy GT resonance found in
(p,n) studies.  Presumably the missing strength is pushed
to higher excitation energies, where it would influence low-energy
neutrino reactions very little.  Thus the relevant issue for us is
to determine how much of the sum rule is exhausted by accessible
strength.  In the case of
$^{208}$Pb, the naive shell model description (closed proton
and neutron major shell at 82 and 126, respectively) predicts
that the $(\bar \nu_e$,e$^+$) direction is completely blocked.
The strength in the $(\nu_e$,e$^-$) direction has been measured 
by forward-angle (p,n) scattering \cite{horen}.  Consistent with the general
trends of GT strength distributions with N-Z, the centroid of the distribution
for this neutron rich nucleus is quite low, just 0.4 MeV above the
position of the IAS.  The resonance is quite narrow and can be reasonably 
fit by
a Gaussian with a full width at half maximum $\Gamma = 2(\ln2)^{1/2}\Delta$
$\sim 4$ MeV and with total strength equivalent to above 46\%
of the Ikeda sum rule \cite{sugarbaker}.  Thus 
\begin{equation}
(g_A^{eff})^2 |M_{\rm GT}(E)|^2 \sim {96.2 \over \Delta \sqrt{\pi}} 
\exp[-(E-E_{\rm GT})^2/\Delta^2],~~^{208}{\rm Pb}(\nu_e,e^-)^{208}{\rm Bi},
\end{equation}
where $E_{GT} \sim$ 17.9 MeV and $\Delta \sim$ 2.4 MeV.
The strength assigned above comes from normalizing the (p,n) cross
section to that for the Fermi transition \cite{sugarbaker},
which is probably the most reliable normalization given the paucity
of strong GT transitions of known strength among heavier nuclei.
However, the ``universal scaling" approach, which depends on the
(p,n)/$\beta$ decay proportionality derived primarily from lighter
nuclei, would reduce the integrated strength in the $^{208}$Bi
peak to 64\% of the value above \cite{sugarbaker}.  Therefore it is not unreasonable
to assign a $\pm$ 50\% uncertainty to this GT resonance estimate.

The light nuclei of interest, $^9$Be, $^{23}$Na, and $^{35}$Cl,
lie in the middle of shells, so consequently both the
$(\nu_e$,e$^-$) and $(\bar \nu_e$,e$^+$) channels are open.
In these cases GT strength distributions are taken from shell model calculations in which
all configurations in the $1p$ or $2s1d$ shells, as appropriate,
are allowed to interact.  This guarantees that the Ikeda sum 
rule is preserved.  The interactions used are Cohen and Kurath
\cite{cohen} and Brown-Wildenthal \cite{bw}.
These calculations, of course, determine both the integrated
GT strength and its distribution.  We use $g_A^{eff} \sim$ 1
to take into account the empirical discrepancy between the
results of such sum-rule-preserving calculations and experimental
estimates of quenching in the region of the GT resonance.

In allowed neutral current neutrino
scattering, the analog of the Fermi operator only contributes to
elastic scattering.  Thus inelastic allowed transitions
are governed by the neutral current GT transition probability
\begin{equation}
|M_{\rm GT}^{\rm NC}|^2 ={1 \over 2J_i+1}
|\langle J_f||\sum_{i=1}^A\sigma(i){\tau_3(i)\over 2}||J_i\rangle|^2.
\end{equation}
This operator is closely connected to the isovector M1 operator,
as the spin contribution to the M1 operator tends to 
dominate because of the large isovector magnetic moment,
$\mu_{\rm V}$ = 4.706.  The distribution of M1 strength in $^{208}$Pb
has been the subject of a great deal of study.  Experimental
searches for the M1 strength \cite{laszewski,kohler} and
theoretical efforts to identify the quenching effects of
correlations \cite{cha,lipparini} has led to a reasonably
consistent picture of the underlying physics.
The simplest closed-shell description attributes the M1 response
to proton $(h_{9/2}) (h_{11/2})^{-1}$ and neutron
$(i_{11/2}) (i_{13/2})^{-1}$ particle-hole excitations.  
The residual interaction mixes these configurations, with the
symmetric combination that saturates the isoscalar response
centered at an excitation energy of about 5.8 MeV,
while the isovector response (the quantity of interest to us)
is centered on a resonance straddling the neutron breakup
threshold at 7.368 MeV.  
The quenching, attributed to more complicated multi-particle-hole correlations, reduces the
naive isovector B(M1) from $\sim \, 50 \,\mu_N^2$ (
nucleon Bohr magnetons squared)
to $\sim  \, 20 \ \mu_N^2$.  Experiment finds 8.8$\mu_N^2$ below the neutron
breakup threshold, and 6.8$\mu_N^2$ immediately above.  Theory
\cite{cha} finds a weak tail of strength at excitation energies
between 10 and 20 MeV of about 0.6$\mu_N^2$.

The integrated isovector B(M1) strength (in units of
$\mu_N^2$) can be related to the neutral current response
\begin{equation}
B(M1) = {3 \mu_{\rm V}^2 \over 4\pi} |M_{\rm GT}^{\rm NC}|^2 \eta^2
\end{equation}
where
\begin{equation}
\eta = 1 + {\langle J_f||\sum_{i=1}^A l(i) \tau_3(i)||J_i \rangle
\over \mu_{\rm V} \langle J_f||\sum_{i=1}^A \sigma(i) \tau_3(i)||J_i \rangle }
\end{equation}
We find $\eta$ = 0.894 using the simple particle-hole description
of the $^{208}$Pb isovector M1 resonance (in effect assuming that
a ratio of orbital and spin matrix elements will not be
greatly changed when correlations responsible for quenching are turned on).
The choices $E_{GT}$ = 7.32 MeV and $\Delta$ = 0.6 MeV yield
a reasonable fit to the measured width of the isovector M1
response and the proper straddling of the neutron breakup
threshold.  So adopting the experimental isovector M1 strength of
(8.8+6.8+0.6) $\mu_N^2$, the distribution of allowed strength
for neutral current neutrino scattering is obtained,
\begin{equation}
(g_A^{eff})^2 |M_{\rm GT}^{\rm NC}(E)|^2 \sim {6.1 \over \Delta \sqrt{\pi}} \exp[-(E-E_{\rm GT})^2/\Delta^2],~~^{208}{\rm Pb}(\nu_i,\nu_f)^{208}{\rm Pb},
\end{equation}
where $E = E_{\nu_i} - E_{\nu_f}$ is the nuclear excitation
energy in $^{208}$Pb.  Approximately 55\% of this distribution lies
below neutron breakup threshold and thus does not contribute to
the spallation.  The corresponding allowed cross section is
\begin{equation}
\sigma(E_{\nu_i}) = 
{G^2_F \over \pi} E_{\nu_f}^2  
(g_A^{\rm eff})^2 |M_{\rm GT}^{\rm NC}|^2.
\end{equation}
  
Average energies of heavy-flavor neutrinos are sufficiently 
high that
odd-parity transitions generated by first-forbidden operators ---
those proportional either to the momentum/energy transfer 
or to nucleon velocities --- must be considered.  
In the case of the simplest nucleus under study, $^9$Be, 
the charged and neutral current responses were evaluated by
including the full momentum transfer dependence of the weak interaction operators,
following Refs. \cite{donnelly,walecka}, and summing
to all 0$\hbar \omega$ and 1$\hbar \omega$ final states.
The 1$\hbar \omega$ shell model space is formed from the 
one-particle-one-hole excitations of the form $1p(1s)^{-1}$ and
$2s1d(1p)^{-1}$; the corresponding cross shell interactions are
the Serber-Yukawa force and the Millener-Kurath interaction \cite{millener}.
As the Slater determinants are formed from harmonic oscillator
basis states, the calculation
is complete for all first-forbidden operators,
which is our main concern.  While high multipolarity operators are
also included in the calculation, the space of final states is not
complete for these.  Nor are these operators significant numerically.
  
As the analogous shell model spaces for the heavier nuclei of interest
become somewhat unwieldy, in these cases we estimate the first
forbidden response in the Goldhaber-Teller model \cite{dubach}.
This model satisfies the Thomas-Reiche-Kuhn (TRK) sum rule for the
E1 response as well as its generalization for L=1 axial responses.
That is, the full supermultiplet of giant resonances is described.
Transition strengths are carried by doorway states
placed in the center of the giant resonance region,
which we identify with the E1 photoabsorption peak
for neutral current reactions.  Note that the
model as implemented here assumes N=Z, which is clearly not the
case for $^{208}$Pb.  However the underlying TRK sum rule is proportional
to $NZ/A=(A/4)\{1-[(N-Z)/A]^2\}$.  Therefore, even for $^{208}$Pb the
total strength prediction, $NZ/A \sim A/4$ is good to 5\%.    
Recently continuum RPA calculations of first-forbidden neutrino
responses were compared to Goldhaber-Teller predictions for 
very neutron rich nuclei \cite{qianr}.  The cross sections agreed to 
better than
40\%.  Thus the expected uncertainties in using this approximation
are not dissimilar to some of those we encountered in our discussions
of the allowed responses.

For $^{23}$Na and $^{35}$Cl, the giant resonance excitation energies,
relative to the parent ground states, were taken to be 19 and 20
MeV, respectively, for both charged and neutral current excitations.
These values are consistent with the observed E1 photoabsorption
peaks.  For neutral current excitations in Pb, we again use the 
E1 photopeak, 14 MeV, to fix the excitation 
energy.  For $^{208}$Pb($\nu_e,e^-)$, the centroid of the spin L=1
strength seen in (p,n) scattering lies about 6.5 MeV above the
isobaric analog state in $^{208}$Bi, corresponding to an excitation
energy of 24.1 MeV relative the ground state of $^{208}$Pb.
Thus we adopt this as the excitation energy.
The strongest first-forbidden contributions to neutrino reactions
are spin modes (0$^-$,1$^-$, and 2$^-$).

We do not use the Goldhaber-Teller model to estimate the 
$^{208}$Pb(${\bar \nu}_e,e^+)$ cross section because, in this direction, the
first-forbidden response in largely blocked: only the
$1h_{11/2}(p) \rightarrow 1i_{11/2}(n)$ transition is allowed
in the naive shell model.  The N $\sim$ Z assumption thus cannot
be used.  However, while we provide no estimate of the cross
section, the almost complete blocking of both the allowed and
first-forbidden response combined with the Coulomb suppression of positron
emission should make this cross section quite small.

The total inelastic cross sections are summarized in Table 1.
Results are shown for ten representative neutrino spectra
and for all of the relevant interactions, so that any 
oscillation scenario can be explored.
The first four, in the absence of oscillations, would be
appropriate for heavy flavor neutrinos, and we believe the 
differences in these spectra are representative of plausible
spectral uncertainties.  The first three of these have
$\eta_{eff}$ = 0, motivated by the Wilson
and Mayle calculations, with a range of average energies
of 30, 25, and 20 MeV.  That is, while 25 MeV might be a 
best guess for the heavy flavor neutrino mean energy, we want to consider the 
consequences of a $\pm$ 20\% uncertainty in average neutrino energy, which
we think in not unreasonable given supernova modeling 
uncertainties.  The fourth case corresponds to a 25 MeV average
energy, but has $\eta_{eff}$ = 3.0, producing a shape more 
similar to the numerical spectrum of Janka.  The last six spectra
all have $\eta$ = 3.0; the first three of these correspond
to average neutrino energies of 19.2, 16, and 12.8 MeV, and
thus are typical of supernova $\bar \nu_e$'s, assuming a 20\%
uncertainty around a best value of 16 MeV.  Similarly, the
last three spectra, with averages energies of 13.2, 11, and
8.8 MeV, are typical of the $\nu_e$'s.

There are some generic features of the cross sections for 
light nuclei in Table 1.
As one would expect, the charged and neutral current cross sections are 
dominated by allowed transitions for lower neutrino temperatures,
with the forbidden contributions becoming increasingly important
as the temperature rises.  For the most energetic spectra,
these two contributions are comparable.
Furthermore, for the highest energies which are typical of heavy flavor
neutrinos, the ratio of the charged current cross section 
to the neutral current cross sections (per flavor)
is in the range of 3 to 5.  Neither of these observations
is particularly welcomed from the experimental viewpoint.
The presence of an appreciable forbidden contribution enhances
the sensitivity of the spectrum-averaged cross section to the
particular shape of the distribution.  Crudely speaking, the forbidden 
cross sections 
contain two extra powers of the neutrino energy.  Therefore it
appears that, in the absence of an independent measurement of the
shape of the energy distribution of the heavy neutrino spectrum, 
plausible spectral
uncertainties could change rate predictions by a factor of three
or more.  The charged to neutral current cross section ratio
is unfortunate because it suggests that the electron and heavy
flavor neutrinos would make, in the most favorable case of a 
hot $\nu_e$ spectrum following an oscillation, comparable contributions 
to total
counting rates. In this case there would be no strong flavor sensitivity.
For example, making the assumption of the same luminosity per 
flavor, a $\nu_e \rightarrow \nu_\tau$ oscillation would 
result in an overall increase in the rate of inelastic 
neutrino scattering events by a factor of $\sim 1.8$
in the case of $^{23}$Na, taking $\nu_e$, $\bar \nu_e$,
and heavy flavor average energies of 11, 16, and 25 MeV, and
assuming $\eta_{eff}$ = 0.0 for the heavy flavor spectrum.
Furthermore we will soon see that most of this enhancement provides no 
neutrons and is thus not detectable. Thus the rate change is comparable to
 the (probably optimistic) 
estimates we made above of cross section uncertainties ($\pm$ 50\%), 
and is dwarfed by the
factor-of-several uncertainties associated with plausible 
spectrum variations.  While our main discussion of these
issues is deferred to the next section, it is already clear 
that tricks will be needed to extract oscillation signals from
neutron spallation yields.

The $^{208}$Pb cross sections require separate discussion, given 
that estimates have already been made by Hargrove \cite{hargrove}.
His allowed neutral current cross section is about a factor of six larger than
ours; a factor of about 1.5 of this appears attributable to his
somewhat less detailed treatment of the M1 strength profile.
The remainder of the discrepancy may be a mistake
in the normalization of his $\beta$ strength function, which
appears to lack the factor of 2 found in Eq. (4).  
(Hargrove also placed all of his strength above the neutron 
threshold, while we noted that in excess of 55\% of the isovector response is
to bound states.  Thus our allowed cross sections for neutron
emission differ by more than an order of magnitude.)  However
Hargrove did not include first-forbidden contributions, which
we find dominate the cross sections for all but the least energetic
spectra.  For example, our $\eta_{eff}$ = 0,
 $\langle E \rangle$ = 25 MeV cross
section is 4.55 $\times$ 10$^{-40} \rm{cm}^{2}$, 89\% of which comes from
first forbidden contributions.  The importance of 
first forbidden contributions in $^{208}$Pb is not
surprising given the dependence of the Thomas-Reiche-Kuhn sum rule 
on N and Z, $\sim$ NZ/A $\sim$ A/4, and the lower energy of the
$^{208}$Pb dipole peak.
This total cross section can be compared to that
of Hargrove, 3.13 $\times 10^{-40}$ cm$^2$.
The end results are not too different, even though most of our
cross section is generated by first forbidden operators
not previously considered.

The first-forbidden contributions to charged current cross sections
are also very important, about twice the allowed contribution
for $\langle E \rangle \sim$ 25 MeV.  Their influence for lower temperatures
is not as great because of the substantially threshold for excited
S=1 L=1 giant resonances.  Making the same comparison as above
to Hargrove, we find our allowed cross sections for $\eta = 0.0$
and $\langle E \rangle$ = 25 MeV are in excellent agreement, 20.3 vs 21.9
in units of $10^{-40}$ cm$^2$.  But our total cross section is
substantial larger, 58.0, due to the giant resonance contributions.
These differences become particularly interesting when we examine
the corresponding spallation cross sections.

The last issue is the probability for producing a signal of one or
more spalled neutrons.  In the case of the lighter
nuclei, unbound states reached by neutrino interactions frequently
decay by competing n, p, or alpha channels.  We have estimated the
neutron emission portion of this cross section by doing Hauser-Feshbach
calculations of the decay probabilities as a function of 
nuclear excitation energy, folding these with the various neutrino 
cross sections $\sigma(E_\nu)$ corresponding to the total cross sections
in Table I.  The resulting neutron emission probabilities
are given in Table II.  Our Hauser-Feshbach calculations are 
reasonably simple in that they employ a nuclear density-of-states
formula that is independent of spin and parity and optical
potentials of the Wood-Saxon type without spin-orbit interactions.
No attempt is made to estimate direct reaction contributions.
Our treatment is identical to that used by Woosley et al. and 
employs the same code and optical model parameterization \cite{woosley}.
One combines the neutron emission probabilities in Table II
with the cross sections in Table I to obtain the needed spectrum-
averaged neutron spallation cross sections.

The case of $^{208}$Pb is simpler because the enormous Coulomb barrier
strongly suppresses charged particle emission.  In the case of neutral
current excitations, the M1 strength is concentrated in a resonance
straddling the neutron emission threshold of 7.37 MeV, as described
previously.  The neutron resonance measurements of Ref. \cite{kohler}
show that neutron emission dominates over gamma decay even immediately
above threshold.  Thus the allowed contribution to single neutron
emission can be calculated by integrating the cross section over the
continuum.  The first forbidden cross section was estimated in the
Goldhaber-Teller model, with the doorway state placed at the peak of
the photoabsorption giant dipole response at $\sim$ 14 MeV.  This
again straddles an important threshold, as two-neutron emission can
occur above 14.1 MeV.

The systematics of two-neutron vs. single neutron emission are well
studied.  For heavy nuclei there is a surprisingly sharp transition
between these two channels occurring typically 2.2 MeV above the
two-neutron threshold \cite{landholt2}.  As this transition is sharp
compared to the breadth of the photoabsorption peak, which has a full
width at half maximum $\Gamma \sim$ 4.3 MeV \cite{danos}, it is a very
reasonable approximation to associate transitions below 16.3 MeV with
single neutron emission, and transitions above this energy with two
neutron emission.

The emission probabilities in Table II were calculated by smearing the
Goldhaber-Teller results over doorway states distributed according to
the measured photoabsorption peak, described as a Gaussian with the
above value of $\Gamma$.  We find that neutral current excitations
almost always lead to single neutron emission.  The two neutron
emission contributions do not exceed 3\%.  The result that neutral
current reactions produce very few multiple neutron events is rather
insensitive to the precise description of the photopeak.  For example,
if the width is increased by a factor of two, the two-neutron emission
probability still remains below 10\%.

We will argue that this conclusion - that neutral currents effects
can be filtered out by observing multiple neutron events - is
quite important for oscillation searches.  It depends on an
assumption, that the spin dipole resonances are located at about
the same place as the photoabsorption giant dipole resonance.

Spallation following the charged current reaction $^{208}{\rm
  Pb}(\nu_e,e^-)^{208}{\rm Bi}$ differs in an important way.
Transitions to states above 6.89 MeV in $^{208}$Bi can emit a neutron,
above 14.98 MeV can emit two neutrons, and above 22.02 MeV can emit
three.  The peak of the Gamow-Teller distribution is at 15.5 MeV.
Thus a small fraction ($\sim$ 10\%) of the allowed charged current
cross section can produce multiple neutrons.  However the L=1
strength, which dominates the heavy flavor neutrino cross section, is
centered at $\sim$ 21.7 MeV, far above the two neutron threshold, and
thus always produces multiple neutrons.

Table II gives the resulting neutron emission probabilities.  In these
calculations, we again attribute all transitions to states above 17.2
MeV in $^{208}$Bi (i.e., 2.2 MeV or more above the two-neutron
threshold) to multiple-neutron decay.  While the single proton
emission channel is also open, the Coulomb barrier provides large
suppression.  Our Hauser-Feshbach calculations yield a very small
ratio of of single proton to single neutron emission throughout the
excitation energy region spanned by the Gamow-Teller and spin-flip
giant resonance peaks.

We repeat for Pb the calculation performed earlier for $^{35}$Cl. That
is, we evaluate rates with and without a $\nu_e \leftrightarrow
\nu_\tau$ oscillation for the canonical temperatures in Table I and
under the assumption of a fixed luminosity per flavor, considering all
spallation events.  One finds that oscillations increase the rate for
all neutron producing events by a factor of $ \sim \, 4$, which is
comparable to the effects of a $\pm$ 20\% change in the heavy neutrino
spectrum temperature.  This is an interesting change, but perhaps not
enough to convince skeptics that the $\nu_\tau$ has a mass.  The
situation is improved relative to $^{35}$Cl because the enhanced
charged current cross sections for this high Z target yield a
favorable ratio of charged to neutral current cross sections.  Thus
the change in the charged current rate due to oscillations, a huge
factor of $ \sim 36$, is discernible despite neutral current
contributions from all other flavors.

But we now see that the situation can be made much, much better.  The
neutral current signal can be all but turned off by counting only
multiple neutron events, while the charged current contribution after
oscillations is only modestly reduced.  That is, the definitive signal
of $\nu_e \leftrightarrow \nu_\tau$ oscillations in a $^{208}$Pb
detector is a dramatic enhancement in multiple neutron events.  A
repetition of the calculation above for multiple neutron events yields
a ratio of multineutron events with oscillations to those without of $
\sim 40$.  In the next section we turn to a more quantitative
exploration of this and other strategies for detecting oscillations.

\section{Strategies for Detecting Flavor Oscillations: Results and Conclusions}
\label{sec:detect}

In this section we will discuss event rates and possible strategies
for LAND, OMNIS, and similar neutron spallation supernova neutrino
observatories.  The calculations presented in the previous section
were performed for specific isotopes of the materials that have been
proposed for these targets.  For example, $^{208}$Pb comprises
slightly more than half of natural lead, while $^{35}$Cl and $^{23}$Na
comprise 75\% and 100\% of natural chlorine and sodium, respectively.
(One of the proposed target material in OMNIS in salt.)  Therefore a
simplification we make is to treat these target materials as being
composed of the principal isotopes.  Given that we are concerned with
neutrino spectrum uncertainties that can change rates by factors of
$\sim$ 3, more detailed modeling is difficult to justify.  In the case
of Pb, the responses are governed by sum rules proportional to N-Z or
NZ/A, quantities that vary little from A=208 to A=206, for example.
For chlorine one anticipates that our charged current allowed cross
sections will be a bit low, given that the ignored isotope $^{37}$Cl
has N-Z=3 and is more neutron rich than $^{35}$Cl.

Interest has been expressed in $^9$Be because its neutron emission
thresholds are so low \cite{goldhaber}.  In some sense it can be
viewed as a neutron target.  Its inclusion is also interesting as a
theory benchmark, since full shell model calculations satisfying both
the allowed and first-forbidden sum rules could be performed.  There
is general consistency among the first-forbidden responses in Table I
for $^9$Be, $^{23}$Na, and $^{35}$Cl, even though the last two were
evaluated in the somewhat schematic Goldhaber-Teller model.

Because of urgent issues such as a cosmologically interesting muon and/or
tauon neutrino mass, proposed supernova neutrino observatories have as
their goal the observation of at least the entire galaxy.  Thus a
typical horizon for such detectors is on the order of the galactic
radius, $\sim$ 10 kpc.  We begin by expressing the neutrino fluence at
earth normed to such a galactic distance.  The total number fluence of
a given neutrino species (e.g., $\nu_e$, ${\bar \nu_e}$, $\nu_\mu$,
etc.) is
\begin{equation}
\label{eq:flux}
\Phi_\nu \approx
2.67 \times 10^{12} {\rm cm}^{-2} 
\left({E_{\rm explosion} \over 3 \times 10^{53} {\rm ergs}} \right)
\left({\rm MeV} \over {\langle E_\nu \rangle} \right)
{1 \over r_{10 \, {\rm kpc}}^2}.
\end{equation}
assuming a total energy in neutrinos of 3 $\times 10^{53}$ ergs, and
an equipartition of energy among the six neutrino species, results
consistent with most transport calculations (see, e.g., Ref.
\cite{wilson}).  The exact distribution of energy among the neutrino
species will be an additional source of error, but considerably
smaller then that associated with the uncertain spectral distribution.
The distance to the supernova, $r_{10 \, {\rm kpc}}$, is given here in
units of 10 kiloparsecs.  As a consequence of the equipartition of
energy, a neutrino species characterized by a lower average energy
will have a higher fluence than one with higher average energy.  All
of our detector event totals will be calculated with this standard
fluence; results for other distances and total explosion energies can
be obtained by appropriately scaling to Eq. \ref{eq:flux}.

In Table III we present the resulting neutron (and multiple neutron
Pb) supernova events, summed over flavor, for one-tonne Pb, NaCl, and
Be targets, given our assumed normalized neutrino fluence of Eq.
\ref{eq:flux} for a standard distance of 10 kpc.  In the case of
neutral current interactions, total inelastic cross sections of these
targets (that is, summed over all subsequent decay channels) are not
very different when quoted per target mass (or per nucleon): values
are within a factor of two of 1.2 $\times 10^{-42}$ cm$^2$ per nucleon
for E$_\nu \sim$ 25 MeV.

This is consistent with naive expectations.  The
forbidden contributions are significant and scale, according to the
Thomas-Reiche-Kuhn sum rule, approximately as A.  Targets are
distinguished, however, by the ease with which they emit neutrons.  In
the case of $^{35}$Cl and $^{23}$Na, the greater phase space for
proton emission tends to dominate over Coulomb effects, leading to
neutral current neutron spallation probabilities of only $\sim$ 10\% .
But $^{208}$Pb and $^9$Be are more favorable cases, the former because
of inhibiting Coulomb barriers and the latter because of an
exceptionally low threshold for neutron emission.  Thus neutron
emission is the dominant decay channel for Pb and Be, producing about
an order of magnitude more signal than in a salt detector of equal
mass.

While the neutron yield is important in efforts to constrain neutrino
masses kinematically, flavor specificity may be more crucial in $\nu_e
- \nu_\tau$ oscillation tests.  That is, does an anomalously hot
$\nu_e$ spectrum produce a distinctive signal in a detector?  A salt
detector, unfortunately, remains problematic.  Such an oscillation
raises the charged current cross section from an insignificant level
to a value comparable to the neutral current cross section summed over
flavors.  But the $(\nu_e,e^-)$ reaction moves one to the proton-rich
side of the parent nucleus, yielding neutron spallation probabilities
of at most a few percent.  The net result is that the
oscillation-induced change in total neutron events is quite modest,
and would be obscured by existing uncertainties in heavy flavor
spectra.  To illustrate this point, in Figure 1 we plot neutron events
with and without oscillations.  In each case there is a band of values
corresponding to the range of spectrum choices used in Tables I and
II, reflecting existing uncertainties in our knowledge of the neutrino
spectra.  As the bands, with and without oscillations, overlap
substantially, it is clear that neutrino spectral uncertainties will
obscure plausible oscillation-induced enhancements of the charged
current events.

The $^9$Be case is somewhat different.  The neutron yields
following $(\nu_e,e^-)$ are exceptionally small, regardless
of oscillations.  The reaction $({\bar \nu_e},e^+)$ has a small
cross section but a high neutron yield per reaction; but even
in the event of antineutrino oscillations, the effect on the
total yield (neutral and charged, summed over flavor) is
about 10\%.  Thus $^9$Be is a relatively clean neutral 
current detector.

This property of a $^9$Be target suggests the possibility of reducing
spectral uncertainties by comparing ratios of rates for different
nuclear targets.  Given that $^9$Be measures the neutral current
response and has perhaps the most easily calculable cross section, it
can be considered a monitor of the heavy flavor temperature: the
neutral current rate is not altered by oscillations.  Thus, by
comparing the event rate in a target with a strong charged current
response to that of $^9$Be, one might hope to remove much of the
uncertainty associated with unknown aspects of the $\nu_\tau$ and
$\nu_\mu$ spectra.  Studies of ratios of events might also prove
helpful if the distance to the supernova were not known.
Superficially this sounds quite attractive as the heavy flavor
spectrum also determines the enhanced charged current response
following oscillations.

To address this issue more quantitatively, we calculated the ratio of
the NaCl events to Be events with and without oscillations.  All
neutron-producing channels are included, and the heavy flavor,
$\nu_e$, and ${\bar \nu_e}$ spectra are allowed to vary over the
ranges in Tables I and II.  The resulting ranges for the ratios, which
are narrower than those of Figure 1, are shown in Figure 2.  While
this strategy clearly has helped in reducing sensitivity to variation
in the spectra, there remain additional uncertainties that affect the
ratio, particularly cross section uncertainties.  The extended cross
section error bars shown in Figure 2 result from combining a $\pm$
50\% uncertainty in the cross section for each target material (Be and
NaCl).  We regard such an uncertainty as an optimistic guess for what
might be achievable, given additional work.  It appears to us that a
definitive claim of oscillations would be difficult to make in a salt
detector, even given a normalizing target such as $^9$Be.

The situation is much improved for a Pb detector.  The first effect
apparent from Table I is the exceptionally strong $(\nu_e,e^-)$ cross
section, a result primarily of the Coulomb enhancement of the cross
section.  As a result, transmutation of $\nu_\tau$'s to $\nu_e$'s
would increase the number of neutron events by a factor of four, as
mentioned previously.  Thus the comparisons in Figures 1 and 2 are
much more favorable.  Even more exciting, of course, in the flavor
specificity provided by multiple neutron events.  The results for
multiple neutron events are shown separately in Figures 1 and 2.  The
enhancement resulting from a complete conversion of $\nu_\tau$'s to
$\nu_e$'s is so large, a factor of 40, that it could not be attributed
to spectral uncertainties.

We conclude that the ability to identify multiple neutron events with
high efficiency in a Pb detector could be of great importance.
Perhaps the most important nuclear structure assumption in the Pb
calculations is the placement of the spin-flip dipole strength for
neutral current excitation at the position of the measured E1
resonance: this leads to the weak neutral current production of
multiple neutrons.  Presumably the location of the dipole spin-flip
strength could checked by spin transfer (p,p') measurements.  If this
strength were located substantially above the E1 giant resonance, our
conclusions would have to be reexamined.

These arguments for finding a signature for neutrino flavor
transformation have been based on the assumption that the energy
spectrum for one type of neutrino species (the electron neutrino) is
completely transformed into that of another (the mu or tau neutrino);
i.e., that neutrino flavor evolution through resonances is adiabatic
over a broad range of neutrino energies that includes high energies.
For the MSW mechanism this will only occur if the density gradient is
sufficiently small in the resonance regions, where the transformation
takes place.  Ignoring neutrino background effects (that is, ignoring
neutrino-neutrino neutral current forward exchange scattering
contributions to the neutrino effective mass \cite{fmws87,qf95}), this
condition on the vacuum mixing angle for adiabatic evolution may be
expressed as \cite{msw},
\begin{equation}
\sin^22\theta {\
\lower-1.2pt\vbox{\hbox{\rlap{$>$}\lower5pt\vbox{\hbox{$\sim$}}}}\ }  
{{4\pi
E_{\nu}}\over{\delta m^2 H}} \approx 6\times {10}^{-2} {\left(
{{E_{\nu}}\over{25\,{\rm MeV}}}\right) } {\left( {{1\,{\rm  
eV^2}}\over{\delta
m^2}}\right) } {\left( {{1\,{\rm km}}\over{H}}\right) },
\end{equation}
where $H \approx {\vert {{1}\over{\rho}} {{d \rho}\over{d r}}
  \vert}^{-1}$ is the density scale height at the resonance position.
Here $\rho$ is the matter density of material at the resonance
position. Again, this expression for $H$ ignores neutrino background
effects. The magnitude of the neutrino-neutrino forward scattering
effects is discussed in Ref.s \cite{fmws87,qf95}.  For a neutrino
energy $E_\nu$ and mixing parameters $\delta m^2$ and $\sin^2 2
\theta$ the resonance density is
\begin{equation}
{\left(\rho Y_e\right)}_{\rm res} \approx 2.6 \times {10}^5\,{\rm g  
\, cm^{-3}}
\left({{\delta m^2}\over 1\,{\rm eV^2}}\right) {\left( {{25\,{\rm
MeV}}\over{E_{\nu}}}\right)} \cos 2\theta ,
\end{equation}
where $Y_e$ is the electron fraction. The relevant densities in the
supernova range from $\rho \approx 10^{12} \, {\rm g} \, {\rm
  cm}^{-3}$ at the surface of the the neutron star to $\rho \approx 10
\, {\rm g} \, {\rm cm}^{-3}$ in the hydrogen envelope.  Therefore, for
small mixing angle, flavor transformation can occur for a range of
$\delta m^2$ of $10^{6} {\rm eV}^2$ to $10^{-5} {\rm eV}^2$.  The most
stringent condition on the mixing angle comes from the outer edges of
the supernova.  Taking densities from Woosley et al. \cite{woosley},
at this location we find a condition on the mixing angle of $\sin^2 2
\theta \gtrsim 10^{-2}$, from Eq. (25).  For higher densities, the
adiabatic condition gives a less stringent limit.

This range of masses and mixings that would be observable in a
supernova includes the popular small-angle MSW solution to the solar
neutrino problem.  This solution has a mass squared difference,
$\delta m^2 \sim 10^{-5} {\rm eV}^2$, see for example \cite{hata}, and
can occur either through transformation between $\nu_e \leftrightarrow
\nu_\tau$, $\nu_e \leftrightarrow \nu_\mu$ or between $\nu_e
\leftrightarrow \nu_s$.  In the first case, a similar crossing would
occur in the supernova at a similar density, $\rho \approx 100-10 \,
{\rm g} \, {\rm cm}^{-3}$.  If on the other hand, if this
transformation occurs by $\nu_e \leftrightarrow \nu_\mu$, then the
seesaw mechanism would predict a $\nu_\tau$ mass of 2 - 100 {\rm eV}
\cite{msw}.  This would necessitate a $\nu_e \leftrightarrow \nu_\tau$
level crossing at high density $\rho \sim 2.6 \times 10^5
\left({{\delta m^2} / 1\,{\rm eV^2}}\right) {\left( {{25\,{\rm MeV}} /
      {E_{\nu}}}\right)}\, {\rm g} \, {\rm cm}^{-3}$, or around $\rho
\approx 10^7 {\rm g} \, {\rm cm}^{-3}$.  There is then an additional
$\nu_e \leftrightarrow \nu_\mu$ crossing at lower density, given a
standard mass hierarchy.  Although this latter scenario presents a
more complicated picture of the neutrino transformations occurring in
the supernova, the effect in terms of neutron count rates seen in the
detector is exactly the same.  Therefore, either of these proposed MSW
solutions to the solar neutrino problem would imply the presence of
matter enhanced neutrino oscillations in the post-core-bounce
supernova.  Finding a signature of matter enhanced neutrino
oscillations in a supernova neutrino detector would provide a 
completely independent check of
this solar solution.  And if the solar neutrino problem proved to have
some other origin, the wider range of mass differences and mixing
angles accessible to supernova neutrino experiments keeps
possibilities open for new physics to emerge there.

Finally, we should stress that our primary focus in this paper has been
on a specific issue, that of finding a signal for flavor
oscillations, including those of the $\nu_\tau$.  
The selection of one target material over another would have
to take into account many other issues, e.g., their comparative
utility in testing the spreading of neutrino arrival times
due to kinematic effects of neutrino masses.  Target materials
will vary in cost, in ease of neutron detection, and in 
ambient backgrounds.  Our efforts have been directed toward
improved event rate estimates and questions of flavor specificity,
in the hope that this information will help experimentalists
make optimal choices.

This work was supported in part by the National Science Foundation
(GF) and the US Department of Energy (GM,WH).

\begin{figure}
\epsfxsize=14cm
\epsfxsize=14cm
\caption{The ranges of expected neutron events given the standard
neutrino fluences discussed in the text, corresponding to 
a supernova at a distance of 10 kpc from earth.
The results are taken from the cross sections and
spallation probabilities of Tables I and II, summed over both
neutral and charge current reactions.  Two ranges are given, without
(left) and with (right) $\nu_\tau$ to $\nu_e$ flavor transformation
(labeled by ft).
The detector materials are Be, NaCl, and Pb, with cross sections
equated to those of the principal isotopes in each case.
A clear signal of oscillations would correspond to a pair of ranges
with no overlap.  Each range is determined from assumed neutrino
spectrum and nuclear physics uncertainties.  The neutrino spectra
are allowed to range over the ($\langle E \rangle$,$\eta$)
values in the Tables,
corresponding to $\pm 20\%$ uncertainties in the canonical  
heavy flavor neutrino,
${\bar \nu_e}$, and $\nu_e$ average energies of 25, 16, and
11 MeV, respectively.  The spectral uncertainties produce the 
inner error bars shown on each range.  These errors have been
further extended by $\pm$ 50\% to indicate possible nuclear
physics uncertainties in our estimated cross sections.
Two sets of results are given for $^{208}$Pb 
corresponding to all neutron-producing events and to all
multiple (denoted by m) neutron events.  
Note the wide separation in the
Pb multiple neutron 
case between the bands with and without oscillations.}
\label{fig1}
\end{figure}

\begin{figure}
\epsfxsize=14cm
\epsfxsize=14cm
\vskip 0.5in
\caption{As in Figure 1, except that ranges for the ratio of
NaCl events to Be events and Pb events to Be events are shown.  The normalized
Pb results are shown for all neutron events and for
multiple neutron events only (labeled by m).  The inner error bars
correspond to the spectral uncertainties, which are reduced
because a ratio has been taken.  The outer error bars show the effects
of cross section uncertainties, which were taken as $\pm$ 50\%
for both the numerators (Pb, NaCl) and denominator (Be) in
taking the ratio of events.}
\label{fig2}
\end{figure}
\centerline{\epsffile{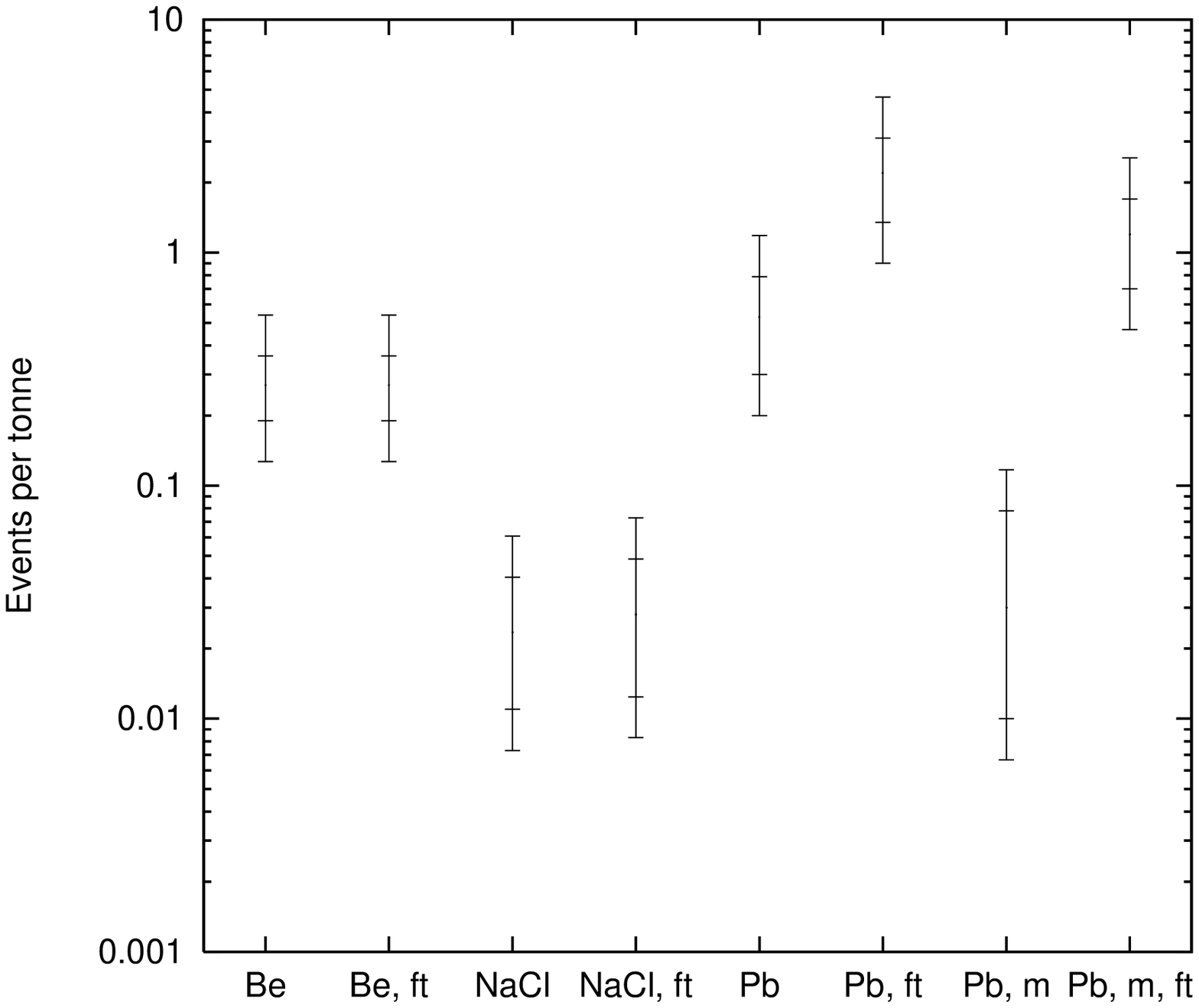}} 
\centerline{\epsffile{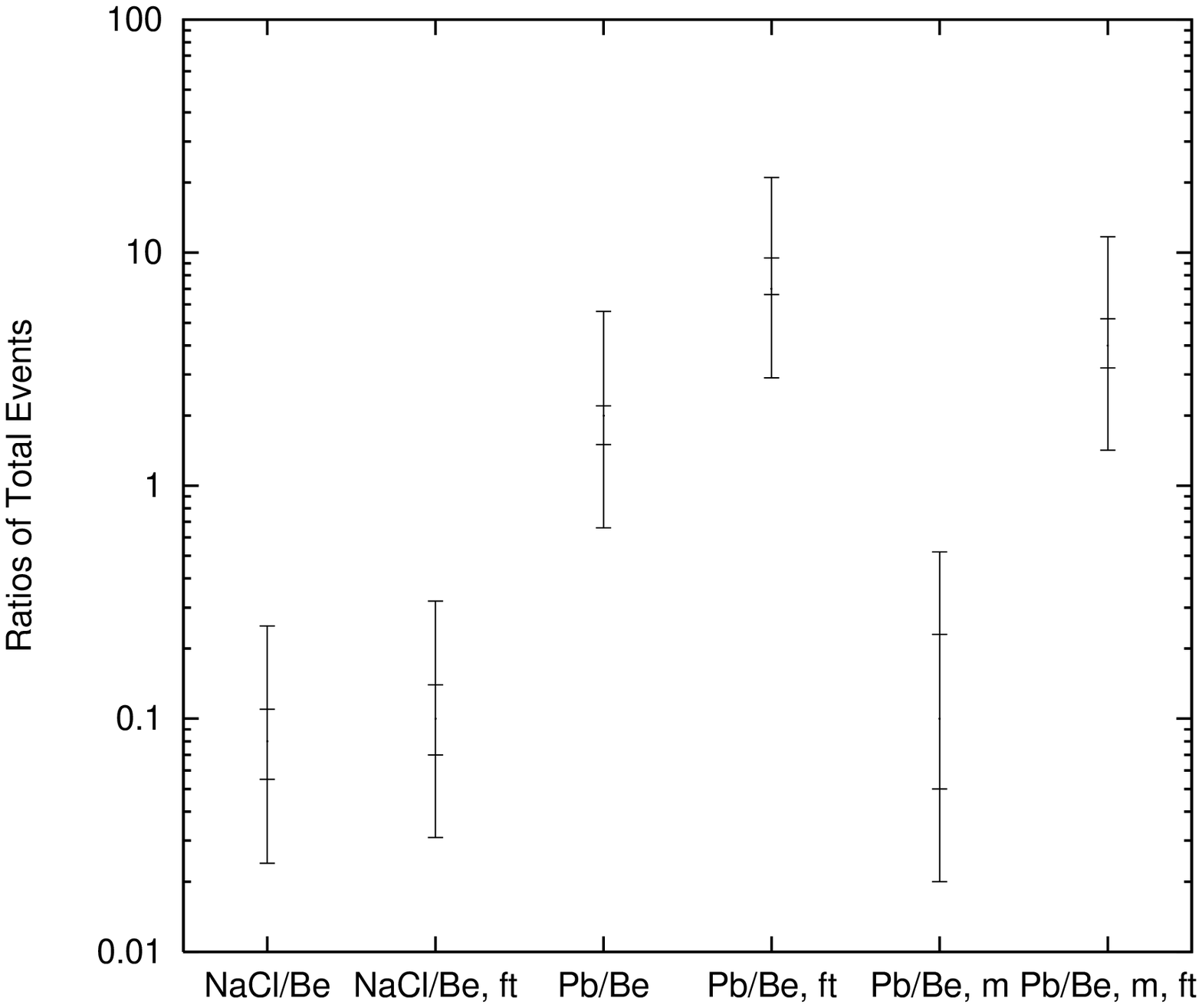}}
\mediumtext
\begin{table}
\caption{Total inelastic neutral current and charge current cross
sections for neutrino reactions on $^{208}$Pb, $^{35}$Cl,
$^{23}$Na and $^9$Be, given in units of 10$^{-40}$ cm$^2$. 
In each case both allowed and
first-forbidden contributions to the cross sections have
been calculated.  The results correspond to normalized neutrino
spectra with a shape defined by the average energy $\langle E \rangle$ and $\eta$,
as discussed in the text.  The first four columns  
describe a range of heavy flavor neutrino spectra centered
around $\langle E \rangle$ = 25 MeV; the next three are appropriate for 
${\bar \nu_e}$s with $\langle E \rangle$ $\sim$ 16 MeV; and the last three 
correspond to $\nu_e$s with $\langle E \rangle \sim$ 11.  
Cross sections are given for each spectrum so that arbitrary
oscillation scenarios can be explored.}
\begin{center}
\begin{tabular}{c c c c c c c c c c c} 
& ${\rm E} = 30$  
& $ {\rm E} = 25 $   
& $ {\rm E} = 20$ 
& $ {\rm E} = 25$
& $ {\rm E} = 19.2$
& $ {\rm E} = 16$
& $ {\rm E} = 12.8$ 
& $ {\rm E} = 13.2$
& $ {\rm E} =  11$
& $ {\rm E} =  8.8$
\\
& $ \eta = 0$  
& $ \eta = 0 $   
& $ \eta = 0$ 
& $ \eta = 3$
& $ \eta = 3$
& $ \eta = 3$
& $ \eta = 3$
& $ \eta = 3$
& $ \eta = 3$
& $ \eta = 3$
 \\
\hline
$^{208}{\rm Pb}\, (\nu,\nu)$ & && &\\
\hline
all & 0.810 &
0.517 & 0.290 & 0.453 & 0.223 & 0.131 & 0.0644 & 0.0714 & 0.0379 &
0.0158 \\
for & 6.423 &
4.032 & 1.996 & 3.388 & 1.288 & 0.527 & 0.157 & 0.188 
& 0.0612 & 0.0125 \\
total  & 7.233 & 4.549 & 2.286 & 3.841 & 1.451 & 0.658 & 0.221 &
0.259 & 0.099 & 0.028 \\ 
\hline
$^{208}{\rm Pb}\, (\bar{\nu},\bar{\nu})$& & &\\
\hline
all &0.810 & 0.517 & 0.290 & 0.453 & 0.223 & 0.131 & 0.0644 & 0.0714 & 0.0379 & .0158  \\
for & 5.220 & 3.308 & 1.664 & 2.825 & 1.046 & 0.457 & 0.139 & 0.166 & 0.055 & 0.0114 \\ 
total & 6.03 & 3.825 & 1.954 & 3.268 & 1.272 & 0.588 & 0.203 & 0.237 & 0.093 & 0.027 \\ 
\hline
$^{208}{\rm Pb}\, (\nu_e,e^{-})$& & &\\
\hline
all  & 34.22 & 20.32 & 10.45 & 17.28 & 7.28 & 3.53 & 1.202 & 1.414 & 0.501 & 0.107 \\ 
for &  61.92 & 37.67 & 17.39 & 30.22 & 9.37 & 3.38 & 0.736 & 0.927 & 0.213 & 0.024 \\
total & 96.14 & 57.99 & 27.84 & 47.50 & 16.65 & 6.91 & 1.938 & 2.341 & 0.714 & 0.131 \\
\hline
$^{35}{\rm Cl}\, (\nu,\nu)$ & && &\\
\hline
all &0.2221 &
0.1488 & 0.0863 & 0.1354 & 0.0671 & 0.0389 & 0.0185 & 0.0206 & 0.0107 
& 0.0044 \\
for &0.2155 & 0.1038 & 0.0370 & 0.0643 & 0.0154 & 0.0049 & 0.0010 
& 0.0013 & 0.0003 & 0.00004 \\
total  & 0.4377 & 0.2527 & 0.1233 & 0.1998 & 0.0825 & 0.0438 
& 0.0195 & 0.0219 &0.0109 & 0.0044 \\ 
\hline
$^{35}{\rm Cl}\, (\bar{\nu},\bar{\nu})$ & && &\\
\hline
all & 0.1820 &  0.1251 &  0.0746 & 0.1162 & 0.0594 & 0.0350 
& 0.0170 & 0.0189 & 0.0099 & 0.0042 \\
for &0.1597 & 0.0792 & 0.0293 & 0.0509 & 0.0127 & 0.0042 & 0.0009 & 0.0011 & 0.0003 & 0.00003 \\
total & 0.3416 & 0.2044 & 0.1039 & 0.1671 & 0.0721 & 0.0392 & 0.0179 & 0.0200 
& 0.0102 & 0.0042 \\
\hline
$^{35}{\rm Cl}\, (\nu_e,e^{-})$ & && &\\
\hline
all & 0.6623 & 0.4229 & 0.2311 & 0.3696 & 0.1695 & 0.0932 & 0.0420 & 0.0471 & 0.0236 & 0.0096\\
for & 0.8306 & 0.3980 & 0.1411 & 0.2455 & 0.0589 & 0.0189 & 0.0039 & 0.0049 & 0.0011 & 0.0001 \\
total & 1.4929 & 0.8209 & 0.3723 & 0.6152 & 0.2284 & 0.1121 & 0.0459 & 0.0519  & 0.0247 & 0.0098 \\
\hline
$^{35}{\rm Cl}\, (\bar{\nu}_e,e^{+})$ & && &\\
\hline
all & 0.0962 & 0.0683 & 0.0432 & 0.0649 & 0.0364 & 0.0233 & 0.0127 & 0.0139 & 0.0081 & 0.0039 \\
for & 0.2229 & 0.1120 & 0.0423 & 0.0735 & 0.0190 & 0.0064 & 0.0014 & 0.0017 & 0.0004 & 0.0001 \\
total & 0.3191 & 0.1804 & 0.0855 & 0.1383 & 0.0554 & 0.0297 & 0.0141 &0.0156 & 0.0085 & 0.0040 \\
\hline
$^{23}{\rm Na}\, (\nu,\nu)$ & && &\\
\hline
all & 0.2071 & 0.1401 & 0.0833 & 0.1282 & 0.0663 & 0.0404 & 0.0211 & 0.0232 & 0.0133 & 0.0066 \\ 
for & 0.1857 & 0.0878 & 0.0309 & 0.0536 & 0.0129 & 0.0042 & 0.0009 & 0.0011 & 0.0003 & 0.00004 \\
total & 0.3928 & 0.2279 & 0.1141 & 0.1818 & 0.0792 & 0.0446 & 0.0220 & 0.0243 & 0.0136 & 0.0066 \\
\hline
$^{23}{\rm Na}\, (\bar{\nu},\bar{\nu}) $ & && &\\
\hline
all & 0.1659 & 0.1153 & 0.0706 & 0.1076 & 0.0575 & 0.0357 & 0.0191 & 0.0209 & 0.0122 & 0.0061 \\
for & 0.1353 & 0.0662 & 0.0242 & 0.0421 & 0.0106 & 0.0035 & 0.0008 & 0.0010 & 0.0002 & 0.00003 \\
total & 0.3012 & 0.1815 & 0.0948 & 0.1497 & 0.0681 & 0.0393 & 0.0199 & 0.0218 & 0.0124 & 0.0062 \\
\hline
$^{23}{\rm Na}\, (\nu_e,e^{-}) $ & && &\\
\hline
all & 0.6992 & 0.4671 & 0.2739 & 0.4231 & 0.2160 & 0.1306 & 0.0677 & 0.0743 & 0.0423 & 0.020 \\
for & 0.6245 & 0.2929 & 0.1022 & 0.1776 & 0.0426 & 0.0139 & 0.0029 & 0.0037 & 0.0009 & 0.0001 \\
total & 1.3237 & 0.7599 & 0.3761 & 0.6007 & 0.2586 & 0.1444 & 0.0706 & 0.0780 & 0.0431 & 0.0205 \\
\hline
$^{23}{\rm Na}\, (\bar{\nu}_e,e^{+})$ & && &\\
\hline
all & 0.0772 & 0.0518 & 0.0303 & 0.0474 & 0.0238 & 0.0138 & 0.0066 & 0.0073 & 0.0037 & 0.001 \\ 
for & 0.2140 & 0.1061 & 0.0397 & 0.0689 & 0.0179 & 0.0061 & 0.0014 & 0.0017 & 0.0004 & 0.0001 \\
total & 0.2913 & 0.1580 & 0.0700 & 0.1163 & 0.0417 & 0.0200 & 0.0079 & 0.0090 & 0.0041 & 0.0015 \\
\hline
$^{9}{\rm Be}\, (\nu,\nu)$ & && &\\
\hline
all &
0.1354 & 0.0933 & 0.0574 & 0.0862 & 0.0473 & 0.0305 & 0.0173 & 0.0188 
& 0.0116 & 0.0063 \\
for & 0.0964 & 0.0428 & 0.0145 & 0.0250 & 0.0062 & 0.0022 & 0.0006 & 0.0007 & 0.0002 & 0.0001 \\ 
total & 0.2317 & 0.1362 & 0.0719 & 0.1112 & 0.0535 & 0.0327 & 0.0179 & 0.0195 & 0.0119 & 0.0063 \\
\hline
$^{9}{\rm Be}\, (\bar{\nu},\bar{\nu})$ & && &\\
\hline
all & 0.1053 & 0.0750 & 0.0478 & 0.0709 & 0.0404 & 0.0267 & 0.0155 & 0.0168 & 0.0106 & 0.0058 \\
for & 0.0659 & 0.0309 & 0.0111 & 0.0191 & 0.0050 & 0.0019 & 0.0005 & 0.0006 & 0.0002 & 0.0001 \\
total & 0.1712 & 0.1059 & 0.0589 & 0.0899 & 0.0455 & 0.0285 & 0.0161 & 0.0174 &  0.0108 & 0.0059 \\
\hline
$^{9}{\rm Be}\, (\nu_e,e^{-})$ & && &\\
\hline
all &
0.7233 & 0.5066 & 0.3202 & 0.4723 & 0.2692 & 0.1796 & 0.1077 & 0.1156 & 0.0754 & 0.0442 \\
for & 0.3268 & 0.1465 & 0.0504 & 0.0866 & 0.0222 & 0.0082 & 0.0023 & 0.0027 & 0.0009 & 0.0002 \\
total & 1.0500 & 0.6531 & 0.3707 & 0.5589 & 0.2914 & 0.1877 & 0.1099 & 0.1184 & 0.0763 & 0.0444 \\
\hline
$^{9}{\rm Be}\, (\bar{\nu}_e,e^{+})$ & && &\\
all & 0.0145 & 0.0084 & 0.0040 & 0.0067 & 0.0025 & 0.0011 & 0.0004 & 0.0004 & 0.0001 & 0.0000 \\
for & 0.0715 & 0.0317 & 0.0103 & 0.0180 & 0.0039 & 0.0012 & 0.0002 & 0.0003 & 0.0001 & 0.0000 \\
total & 0.0860 &0.0401 & 0.0143 &0.0247 & 0.0064 & 0.0023 & 0.0006 & 0.0007 & 0.0002 & 0.0000 \\
\hline
\end{tabular}
\end{center}
\label{tab1}
\end{table}
  
\mediumtext
\begin{table}
\caption{Neutron spallation probabilities for allowed, forbidden,
and all neutrino induced transitions in Be, Na, Cl, and Pb.
The calculations are Hauser-Feshbach type, except in the case
of Pb, as discussed in the text.  The Pb results are given 
separately for single and multiple neutron spallation.}
\begin{center}
\begin{tabular}{c c c c c c c c c c c} 
 E & 30  
& $25 $   
& $ 20$ 
& $25$
& $19.2$
& $16$
& $12.8$ 
& $13.2$
& $11$
& $8.8$
\\
$ \eta$ & $0$  
& $0 $   
& $0$ 
& $3$
& $3$
& $3$
& $3$
& $3$
& $3$
& $3$
 \\
\hline
$^{208}{\rm Pb}\, (\nu,\nu)$ & 1n&& &\\
\hline
allowed & 0.443 & 0.441 & 0.438 & 0.440 & 0.435 & 0.430 & 0.422 & 0.423 & 0.415& 0.403\\
forbidden  & 0.969 & 0.970 & 0.972 & 0.972 & 0.975 & 0.978 & 0.982 & 0.981 & 0.985 & 0.992 \\
\hline
$^{208}{\rm Pb}\, (\bar{\nu},\bar{\nu})$ & 1n & &\\
\hline
allowed & 0.443 & 0.441 & 0.438 & 0.440 & 0.435 & 0.430 & 0.422 & 0.423 & 0.415& 0.403  \\
forbidden & 0.969 & 0.970 & 0.972 & 0.972 & 0.975 & 0.978 & 0.982 & 0.981 & 0.985 & 0.992\\  
\hline
$^{208}{\rm Pb}\, (\nu_e,e^{-})$& 1n& &\\
\hline
allowed  & 0.904 & 0.908 & 0.914 & 0.912 & 0.922 & 0.931 & 0.942 & 0.940 & 0.950 & 0.962  \\ 
forbidden &  0.0 & 0.0 & 0.0  & 0.0 & 0.0 & 0.0 & 0.0 & 0.0 & 0.0 & 0.0 \\
\hline
$^{208}{\rm Pb}\, (\nu,\nu)$ & 2n && &\\
\hline
allowed & 0.0 & 0.0 & 0.0  & 0.0 & 0.0 & 0.0 & 0.0 & 0.0 & 0.0 & 0.0\\
forbidden & 0.031 & 0.030 & 0.028 & 0.028 & 0.025 & 0.022 & 0.018 & 0.019 & 0.015 &0.008  \\
\hline
$^{208}{\rm Pb}\, (\bar{\nu},\bar{\nu})$& 2n & &\\
\hline
allowed & 0.0 & 0.0 & 0.0  & 0.0 & 0.0 & 0.0 & 0.0 & 0.0 & 0.0 & 0.0\\
forbidden & 0.031 & 0.030 & 0.028 & 0.028 & 0.025 & 0.022 & 0.018 & 0.019 & 0.015 &0.008  \\  
\hline
$^{208}{\rm Pb}\, (\nu_e,e^{-}) $& 2n& &\\
\hline
allowed & 0.096 & 0.092 & 0.086 & 0.088 & 0.078 & 0.069 & 0.058 & 0.060 
& 0.050 & 0.038\\
forbidden & 1. 0 & 1. 0 & 1. 0 & 1. 0 & 1. 0 & 1. 0 & 1. 0 & 1. 0 & 1. 0 & 1. 0 \\
\hline
$^{35}{\rm Cl}\, (\nu,\nu)$ & && &\\
\hline
allowed &  0.0032 & 0.0029 & 0.0025 & 0.0026 & 0.0021 & 0.0017 & 0.0012 & 0.0012 & 0.0009 & 0.0005\\
forbidden & 0.0917 & 0.0917 & 0.0917 & 0.0917 & 0.0917 & 0.0917 & 0.0917 & 0.0917 & 0.0917 & 0.0917 \\
total  & 0.0468 & 0.0394 & 0.0292 & 0.0313 & 0.0188 & 0.0118 & 0.0058 & 0.0064 & 0.0033 & 0.0012\\ 
\hline
$^{35}{\rm Cl}\, (\bar{\nu},\bar{\nu})$ & && &\\
\hline
allowed &  0.0032 & 0.0029 & 0.0025 & 0.0026 & 0.0021 & 0.0017 & 0.0012 & 0.0012 & 0.0008 & 0.0005\\
forbidden & 0.0917 & 0.0917 & 0.0917 & 0.0917 & 0.0917 & 0.0917 & 0.0917 & 0.0917 & 0.0917 & 0.0917\\
total &  0.0446 & 0.0373 & 0.0276 & 0.0298 & 0.0179 & 0.0112 & 0.0056 & 0.0062 & 0.0031 & 0.001\\
\hline
$^{35}{\rm Cl}\, (\nu_e,e^{-})$ & && &\\
\hline
allowed &
0.0013 & 0.0011 &0.0009 & 0.0010 & 0.0007 & 0.0005 & 0.0003 & 0.0003 & 0.0002 & 0.000\\
forbidden & 0.0152 & 0.0152 & 0.0152 & 0.0152 & 0.0152 & 0.0152 & 0.0152 & 0.0152 & 0.0152 & 0.0152 \\
total & 0.0090 & 0.0079 & 0.0063 & 0.0066 & 0.0044 & 0.0030 & 0.0015 & 0.0017 & 0.0008 & 0.0003 \\
\hline
$^{35}{\rm Cl}\, (\bar{\nu}_e,e^{+})$ & && &\\
\hline
allowed & 0.4468 & 0.4346 & 0.4150 & 0.4278 & 0.3988 & 0.3723 & 0.3311 & 0.3374 & 0.2968 & 0.2368  \\
forbidden & 0.9046 & 0.9046 & 0.9046 & 0.9046 & 0.9046 & 0.9046 & 0.9046 &  0.9046 & 0.9046 &  0.9046  \\
total & 0.7666 & 0.7266 & 0.6571 & 0.6810 & 0.5720 & 0.4867 & 0.3867 & 0.3998 & 0.3262 & 0.245 \\
\hline
$^{23}{\rm Na}\, (\nu,\nu)$ & && &\\
\hline
allowed & 0.0478 & 0.0419 & 0.0344 & 0.0375 & 0.0277 & 0.0208 & 0.0131 & 0.0141 & 0.0087 & 0.0040 \\ 
forbidden & 0.3058 & 0.3058 & 0.3058 & 0.3058 & 0.3058 & 0.3058 & 0.3058 & 0.3058 & 0.3058 & 0.3058 \\
total & 0.1698 & 0.1436 & 0.1078 & 0.1166 & 0.0729 & 0.0476 & 0.0247 & 0.0273 & 0.0144 & 0.0055 \\
\hline
$^{23}{\rm Na}\, (\bar{\nu},\bar{\nu}) $ & && &\\
\hline
allowed & 0.0487 & 0.0429 & 0.0352 & 0.0386 & 0.0284 & 0.0213 & 0.0133 & 0.0143 & 0.0088 & 0.004\\
forbidden & 0.3058  & 0.3058 & 0.3058 & 0.3058 & 0.3058 & 0.3058 & 0.3058 & 0.3058 & 0.3058 & 0.3058  \\
total & 0.1642 & 0.1388 & 0.1044 & 0.1138 & 0.0716 & 0.0470 & 0.0246 & 0.0271 & 0.0144 & 0.005 \\
\hline
$^{23}{\rm Na}\, (\nu_e,e^{-}) $ & && &\\
\hline
allowed & 0.0041 & 0.0032 & 0.0022 & 0.0025 & 0.0014 & 0.0009 & 0.0004 & 0.0004 & 0.0002 & 0.0001\\
forbidden & 0.0936 & 0.0936 & 0.0936 & 0.0936 & 0.0936 & 0.0936 & 0.0936 & 0.0936 & 0.0936 & 0.0936 \\
total & 0.0463 & 0.0380 & 0.0271 & 0.0294 & 0.0166 & 0.0098 & 0.0043 & 0.0048 & 0.0021 &  0.0006 \\
\hline
$^{23}{\rm Na}\, (\bar{\nu}_e,e^{+})$ & && &\\
\hline
allowed & 0.3561 & 0.3449 & 0.3265 & 0.3362 & 0.3075 & 0.2820 & 0.2437 & 0.2495 & 0.2136 & 0.1650 \\ 
forbidden & 0.5822 & 0.5822 & 0.5822 & 0.5822 & 0.5822 & 0.5822 & 0.5822 & 0.5822 & 0.5822 & 0.5822 \\
total & 0.5222 & 0.5043 & 0.4716 & 0.4819 & 0.4255 & 0.3741 & 0.3019 & 0.3123 & 0.2514 & 0.1810 \\
\hline
$^{9}{\rm Be}\, (\nu,\nu)$ & && &\\
\hline
allowed &
0.7360 & 0.7444 & 0.7545 & 0.7500 & 0.7626 & 0.7716 & 0.7826 & 0.7811 & 0.7896 & 0.7990 \\
forbidden & 0.5208 & 0.5191 & 0.5150 & 0.5155 & 0.5071 & 0.4997 & 0.4904 & 0.4916 & 0.4849 & 0.4799 \\ 
total & 0.6465 & 0.6735 & 0.7063 & 0.6973 & 0.7330 & 0.7531 & 0.7728 & 0.7704 & 0.7835 & 0.7961 \\ 
\hline
$^{9}{\rm Be}\, (\bar{\nu},\bar{\nu})$ & && &\\
\hline
allowed & 
0.7350 & 0.7431 & 0.7533 & 0.7482 & 0.7612 & 0.7704 & 0.7816 & 0.7801 & 0.7889 & 0.7984\\
forbidden & 0.5267 & 0.5245 & 0.5198 & 0.5204 & 0.5117 & 0.5042 & 0.4949 & 0.4961 & 0.4896 & 0.4853 \\
total & 0.6548 & 0.6794 & 0.7093 & 0.6999 & 0.7335 & 0.7530 & 0.7724 & 0.7700 & 0.7831 & 0.7956 \\
\hline
$^{9}{\rm Be}\, (\nu_e,e^{-})$ & && &\\
\hline
allowed &0.0000 & 0.0000 & 0.0000 & 0.0000 & 0.0000 & 0.0000 & 0.0000 & 0.0000 & 0.0000 & 0.0000 \\
forbidden & 0.0068 & 0.0055 & 0.0040 & 0.0040 & 0.0023 & 0.0014 & 0.0007 & 0.0007 & 0.0003 & 0.0001 \\
total & 0.0021 & 0.0012 & 0.0005 & 0.0006 & 0.0002 & 0.0001 & 0.0000 & 0.0000 &0.0000 & 0.0000 \\
\hline
$^{9}{\rm Be}\, (\bar{\nu}_e,e^{+})$ & && &\\
allowed & 0.6599 & 0.6279 & 0.5832 & 0.5932 & 0.5263 & 0.4714 & 0.3943 & 0.4056 & 0.3368 & 0.2481 \\
forbidden & 0.9435 & 0.9560 & 0.9698 & 0.9700 & 0.9831 & 0.9895 & 0.9947 & 0.9942 & 0.9969 &0.9987 \\
total & 0.8958 & 0.8873 & 0.8618 & 0.8673 & 0.8050 & 0.7367 & 0.6242 & 0.6415 & 0.5323 & 0.3851 \\
\hline
\end{tabular}
\end{center}
\label{tab2}
\end{table}
  
\mediumtext
\begin{table}
\caption{The total number of neutron events for one tonne (10$^3$ kg) Pb, 
NaCl, and Be targets, given a neutrino fluence corresponding
to 5 $\times 10^{52}$ ergs per neutrino type ($\nu_e$,${\bar \nu_e}$,
etc.), a supernova distance of 10 kpc, and average neutrino
energies as shown.  Results are shown separately for allowed
and forbidden contributions and, in the case of Pb, for 
single and multiple neutron events.}
\begin{center}
\begin{tabular}{c c c c c} 
& $\nu_\mu + \bar{\nu}_\mu + \nu_\tau + \bar{\nu}_\tau$  
& $ \nu_e $   
& $ \bar{\nu}_e$ 
& Total
 \\
\hline
$^{208}{\rm Pb} $ & & & &\\
\hline
1 n & 0.439 & 0.038 & 0.024 & 0.50 \\
2 n & 0.013 & 0.017 & 0.0005 & 0.030 \\
1n, $\nu_e \leftrightarrow \nu_\mu$ or $\nu_e \leftrightarrow \nu_\tau$ & 0.325& 0.684 & 0.024 & 1.0 \\
2n, $\nu_e \leftrightarrow \nu_\mu$ or $\nu_e \leftrightarrow \nu_\tau$ & 0.010 & 1.20 & 0.005 & 1.2 \\  
\hline
$^{35}{\rm Cl}$&&&&\\
\hline
all n & 0.0064 & 2.3 $\times 10^{-5}$ & 0.0042 & 0.011 \\
all n, $\nu_e \leftrightarrow \nu_\mu$ or $\nu_e \leftrightarrow \nu_\tau$  & 0.0046 & 0.0030 & 0.0042 & 0.012 \\
\hline
$^{23}{\rm Na}$&&&&\\
\hline
all n & 0.0318 & 1.8 $\times 10^{-4}$ & 0.0040 & 0.036 \\
all n, $\nu_e \leftrightarrow \nu_\mu$ or $\nu_e \leftrightarrow \nu_\tau$ & 0.0229 & 0.0169 & 0.0040 & 0.044 \\
\hline
$^{9}{\rm Be}$&&&&\\
\hline
all n & 0.229 & 0.0148 & 0.026 & 0.27 \\
all n, $\nu_e \leftrightarrow \nu_\mu$ or $\nu_e \leftrightarrow \nu_\tau$  & 0.180 & 0.0648 & 0.026 & 0.27 \\
\hline
\hline
\end{tabular}
\end{center}
\label{tab3}
\end{table}

\begin{references}
\bibitem{cline91} D. B. Cline {\em et. al. \/}, Astrophysical Letters
  and Communications, {\bf 27}, 403, 1990.  
\bibitem{cline94} D. B. Cline {\em et. al. \/}, Phys. Rev. D 
 {\bf 50}, 720 (1994), P. F. Smith, Astroparticle Physics, {\bf 8} 27 (1997).  
\bibitem{hargrove} C.
  K. Hargrove {\em et. al. \/}, Astroparticle Physics, {\bf 5}, 183
  (1996).  
\bibitem{balantekin} B. Balantekin, Proceedings of the
  Jorge Andre Swieca Summer School, Campos de Jordao, Sao Paulo, in
  press, (1997).  
\bibitem{lim}C. S. Lim and W. J. Marciano, Phys.
  Rev. D. {\bf 37}, 1368 (1988); E. Akhmedov, Phys. Lett. {\bf 213},
  64 (1988).  
\bibitem{msw} S. P. Mikheyev and A. Yu. Smirnov, Sov. J.
  Nucl. Phys. {\bf 24}, 913 (1985); H. A. Bethe, Phys Rev. Lett. {\bf
    56}, 1305 (1986); W. C. Haxton, Phys. Rev. Lett. {\bf 57}, 1271
  (1986); S. J. Parke, Phys. Rev. Lett. {\bf 57}, 1275 (1986).
\bibitem{solar} J. N. Bahcall, Neutrino Astrophysics, (Cambridge
  University Press, Cambridge 1989); for a recent review see W. C.
  Haxton, Ann. Rev.  Astron. Astrophys. {\bf 33}, 459 (1995).
\bibitem{seesaw} M. Gell-Mann, P. Ramond, and R. Slansky, in
  Supergravity, Proceedings of the Workshop, Stoney Brook, New York,
  1979, edited by P. Van Nieuwenhuizen and D. Z. Freedman
  (North-Holland, Amsterdam, 1979); T. Yanagida, in Proceedings of the
  Workshop on Unified Theory and Baryon Number of the Universe,
  Tsukuba, Japan, 1979, edited by A. Sawada and A. Sugamoto (KEK
  Report No. 79-18, Tsukuba, Japan, 1979).  
\bibitem{babu} K. S. Babu
  and R. N. Mohapatra, Phys. Rev. Lett.  {\bf 70}, 2845 (1993); S. A.
  Bludman, D. C. Kennedy, and P. G. Langacker, Phys. Rev. D {\bf 45},
  1810 (1992); S. Dimopoulos, L. J. Hall, and S. Raby, Phys. Rev. D
  {\bf 47}, R3697 (1993).  
\bibitem{qian} Y.-Z. Qian {\em et al.},
  Phys. Rev. Lett {\bf 71}, 1965 (1993).  
\bibitem{fuller92}G. M.
  Fuller, R. W. Mayle, B. S. Meyer, and J. R. Wilson, Astrophys. J.
  {\bf 389}, 517 (1992).  
\bibitem{lsnd} C. Athanassopoulus {\em et al.}, Phys. Rev. Lett.  
{\bf 75}, 2650 (1995); J. E. Hill, Phys Rev. Lett. {\bf 75}, 2650 (1995).  
\bibitem{nomad} J. M. Gaillard, Nucl. Phys. B {\bf 51}, 237 (1996).  
\bibitem{chorus} K. Kodama, Nucl. Phys B.  {\bf 51}, 232 (1996).  
\bibitem{karmen} B. Armbruster {\em et al.}, Nucl. Phys. B 
{\bf 38}, 235 (1995).  
\bibitem{atmos}
  T. Kajita, talk presented at Neutrino '98 (Takayama, Japan), to be
  published.  
\bibitem{cf} P. F. Harrison, D. H. Perkins, W. G. Scott,
Phys Lett. B {\bf 349}, 137 (1995), C. Y. Cardall and G. M. Fuller, Phys. 
Rev. D{\bf 53}, 4421 (1996), G. L. Fogli, E. Lisi, D. Montanino, Phys. Rev 
D{\bf 54}, 2048 (1996), A. Acker and S. Packvasa, Phys. Lett. B {\bf 397}, 
209 (1997)
\bibitem{mayle} R. W. Mayle and J. R. Wilson, unpublished.
\bibitem{totani} T. Totani, Phys. Rev. Lett. {\bf 80}, 2039 (1998).
\bibitem{sk} M. Takita, in {\it Frontiers of Neutrino Astrophysics},
ed. Y. Suzuki and K. Nakamura (Univ. Acad. Press, Tokyo), p. 135.
\bibitem{burrows} A. Burrows, D. Klein, and R. Gandhi,
Phys. Rev. D {\bf 45}, 3361 (1992).
\bibitem{SNO}G. T. Ewan et al., Queen's Univ. Report SNO-87-12 (1987);
SNO collaboration, Hyperfine Interactions {\bf 130}, 199 (1996).
\bibitem{haxton16o} W. C. Haxton, Phys. Rev. D {\bf 36}, 2283 (1987);
Y.-Z. Qian and G. M. Fuller, Phys. Rev. D
{\bf 49}, 1762 (1994).
\bibitem{minakata} H. Minakata, private communication.
\bibitem{caltech} K. Langanke, P. Vogel, and E. Kolbe, Phys. Rev.
Lett. {\bf 76}, 2629 (1997).
\bibitem{beacom} J. F. Beacom and P. Vogel, Phys .Rev. D58 (1998),
J. F. Beacom, P. Vogel, Phys. Rev. D, submitted (1998). 
\bibitem{wilson} D. S. Miller, J. R. Wilson, and R. W. Mayle,
Astrophys. J., {\bf 415}, 278 (1993).
\bibitem{janka} H.-T. Janka and W. Hillebrant, Astron. Astrophys.,
{\bf 224}, 49 (1989).
\bibitem{behrends} H. Behrends and J. Janecke,
{\it Numerical Tables for Beta Decay and Electron Capture},
Landolt-Bornstein, vol. 4 (Springer-Verlag, Berlin, 1969).
\bibitem{brown} B. A. Brown and B. H. Wildenthal, Phys. Rev. C
{\bf 28}, 2397 (1983).
\bibitem{horen} D. J. Horen {\em et al.}, Phys. Lett.,
{\bf 95B}, 27; C. Gaarde et al., Nucl. Phys. A{\bf 369}, 258 (1981).
\bibitem{sugarbaker} E. Sugarbaker, talk presented at the Workshop on GT and
Neutrino Cross Sections, Univ. Pennsylvania, April, 1993.
\bibitem{cohen} S. Cohen and D. Kurath, Nucl. Phys. {\bf 73}, 1 (1965).
\bibitem{bw} B. H. Wildenthal, Prog. Part. Nucl. Phys. {\bf 11}, 5 (1984).
\bibitem{laszewski} R. M. Laszewski, R. Alarcon, D. S. Dale, and S. D. Hoblit,
Phys. Rev. Lett. {\bf 61}, 1710 (1988).
\bibitem{kohler} R. Kohler et al., Phys. Rev. C{\bf 35}, 1646 (1987).
\bibitem{cha} D. Cha, B. Schwesinger, J. Wambach, and J. Speth,
Nucl. Phys. A{\bf 430}, 321 (1984).
\bibitem{lipparini} E. Lipparini and A. Richter, Phys. Lett. {\bf 144}B, 13 (1984).
\bibitem{donnelly} T. W. Donnelly and W. C. Haxton, Atomic Data and
Nuclear Tables {\bf 23}, 103 (1979).
\bibitem{walecka} J. D. Walecka, in {\it Muon Physics}, ed. 
V. W. Hughes and C. S. Wu (Academic Press, New York, 1975),
Vol. 2, p. 113.
\bibitem{millener} D. J. Millener and D. Kurath, Nucl. Phys. A{\bf 255},
315 (1975).
\bibitem{dubach} T. W. Donnelly, J. Dubach, and W. C. Haxton, Nucl. Phys.
A{\bf 251}, 353 (1975).
\bibitem{qianr} Y.-Z. Qian, W. C. Haxton, K. Langanke, and P. Vogel,
Phys. Rev. C{\bf 55}, 1532 (1997).
\bibitem{woosley} S. E. Woosley, R. D. Hoffman, W. C. Haxton,
Astrophys. J. {\bf 356}, 272, (1990)
\bibitem{landholt2} K. A. Keller, J. Lange, H. Munzel, and G. Pfennig,
{\it Q Values and Excitation Functions of Nuclear Reactions},
Landolt-Bornstein, vol. 5 (Springer-Verlag, Berlin, 1973).
\bibitem{danos} M. Danos and E. G. Fuller, Ann. Rev. Nucl. Science
{\bf 15}, 29 (1965).
\bibitem{goldhaber}M. Goldhaber, private communication.
\bibitem{fmws87} G. M. Fuller, R. W. Mayle, J. R. Wilson, and D. N.  
Schramm, Astrophys. J. {\bf 322}, 795 (1987).
\bibitem{qf95} Y.-Z. Qian and G. M. Fuller, Phys. Rev. D{\bf 51},  
1479 (1995).
\bibitem{hata} N. Hata and P. Langacker, Phys. Rev. D {\bf 56}, 6107 (1997).
\end{references}
\end{document}